\documentclass{article}

\usepackage{amsmath,amssymb,amsfonts}
\usepackage[a4paper]{geometry}
\usepackage{subcaption}
\usepackage{caption}
\usepackage{pifont}
\usepackage{eqparbox}
\usepackage{url} 
\usepackage{multirow}
\usepackage{textcomp}
\usepackage{graphicx}
\usepackage{bm}
\usepackage{dblfloatfix}
\usepackage{anyfontsize}
\usepackage[square,numbers, sort&compress]{natbib}
\usepackage[colorlinks=true, allcolors=blue, urlcolor=black]{hyperref} 
\newgeometry{
    left=2.2cm,    
    right=2.2cm    
}
\usepackage{hyperref}
\hypersetup{
    colorlinks,
    citecolor=black,
    filecolor=black,
    linkcolor=black,
    urlcolor=black
}
\usepackage[affil-it]{authblk}

\begin{document}

\title{Day-Ahead Transmission Grid Topology Optimization Considering Renewable Energy Sources' Uncertainty}

\author[1,2]{Giacomo Bastianel}
\author[1,2]{Dirk {Van Hertem}}
\author[1,2]{Hakan Ergun}
\author[3]{Line Roald}

\affil[1]{{KU Leuven - Electa research group, Leuven, Belgium}}
\affil[2]{{Etch-EnergyVille, Genk, Belgium}}
\affil[3]{{University of Wisconsin-Madison, Madison, United States of America}}

\date{}
\maketitle

\begin{abstract}
The increasing penetration of renewable energy sources (RES) introduces significant uncertainty in power system operations. At the same time, the existing transmission grid is often already congested, and urgently needed reinforcements are frequently delayed due to technical, economic, or societal constraints. To address these challenges, adjusting the grid topology based on congestion patterns is considered a non-costly remedy to guarantee a more efficient power transmission. Based on this idea, this paper proposes a multistep stochastic grid topology optimization model that combines optimal transmission switching and busbar splitting for both AC and hybrid AC/DC grids. The methodology incorporates RES forecast uncertainty through a scenario-based stochastic optimization approach, using real offshore wind data and K-means clustering to generate representative forecast error scenarios. The proposed model includes several formulations to be compared with a plain optimal power flow (OPF) model: hourly optimizing the topology, one topology for a given time horizon (24 hours), or a limited number of switching actions over a given time horizon. The grid topology optimization model is formulated as a Mixed-Integer Quadratic Convex Problem, optimized based on the day-ahead (D-1) RES forecast and validated for AC-feasibility via a full AC-OPF formulation. Based on the generation setpoints of the feasibility check, a redispatch simulation based on the measured (D) RES realization is then computed. The methodology is tested on a modified version of an AC 30-bus test case and a hybrid AC/DC 50-bus test case, for a 24-hours (30-bus test case) and a 14-days (both test cases) time series. The results highlight the economic benefits brought by grid topology optimization for congested test cases with high penetration of RES. In addition, the results demonstrate that accounting for RES uncertainty with at least 6 to 8 scenarios leads to lower or comparable total (generation and redispatch) costs with respect to deterministic day-ahead forecasts, even when limiting the frequency of topological actions.
\end{abstract}




\newcommand{\acnodes}{\mathcal{I}}
\newcommand{\acbranches}{\mathcal{L}}
\newcommand{\acswitches}{\mathcal{SW}^{ac}}
\newcommand{\actopology}{\mathcal{T}^{ac}}
\newcommand{\actopologyrev}{\mathcal{T}^{ac, rev}}
\newcommand{\acswitchtopology}{\mathcal{T}^{\text{sw,ac}}}
\newcommand{\acswitchtopologyrev}{\mathcal{T}^{\text{sw}^{\text{ac, rev}}}}
\newcommand{\acZILtopology}{\mathcal{T}^{\text{ZIL,ac}}}


\newcommand{\nodevoltage}{V_i}
\newcommand{\acbranchflow}{S_{lij}}



\newcommand{\dcnodes}{\mathcal{E}}
\newcommand{\dcbranches}{\mathcal{D}}
\newcommand{\dcswitches}{\mathcal{SW}^{dc}}
\newcommand{\dctopology}{\mathcal{T}^{dc}}
\newcommand{\dctopologyrev}{\mathcal{T}^{dc, rev}}
\newcommand{\dcswitchtopology}{\mathcal{T}^{\text{sw,dc}}}
\newcommand{\dcswitchtopologyrev}{\mathcal{T}^{\text{sw}^{\text{dc}, rev}}}
\newcommand{\dcZILtopology}{\mathcal{T}^{\text{ZIL,dc}}}

\newcommand{\acnodesnew}{\mathcal{I'}}
\newcommand{\acZIL}{\mathcal{S}}
\newcommand{\dcnodesnew}{\mathcal{E'}}
\newcommand{\dcZIL}{\mathcal{Q}}

\newcommand{\timesteps}{T}

\newcommand{\scenarios}{\mathcal{W}}


\newcommand{\dcbranchflow}{P_{def}}



\newcommand{\acdcconverters}{\mathcal{C}}

\newcommand{\convertertopology}{\mathcal{T}^{\text{cv}}}

\newcommand{\generators}{ \mathcal{G}}

\newcommand{\loads}{\mathcal{M}}

\newcommand{\dcgenerators}{\mathcal{G}^{\text{dc}}}

\newcommand{\dcloads}{\mathcal{M}^{\text{dc}}}

\newcommand{\genconn}{\mathcal{T}^{\text{gen}}}
\newcommand{\dcgenconn}{\mathcal{T}^{\text{gen,dc}}}

\newcommand{\acloadconn}{\mathcal{T}^{\text{load}}}
\newcommand{\dcloadconn}{\mathcal{T}^{\text{load, dc}}}


\newcommand{\genpower}{ S^g_k }
\newcommand{\acloadpower}{ S^m_k }
\newcommand{\dcloadpower}{ P^m_k }
\newcommand{\converteracpower}{ S^c_l }
\newcommand{\converterdcpower}{ P^{c, dc}_l }


\newcounter{model1} \setcounter{model1}{0}
\newcounter{model2} \setcounter{model2}{0}
\newcounter{model3} \setcounter{model3}{0}
\newcounter{model4} \setcounter{model4}{0}
\newcounter{model5} \setcounter{model5}{0}
\newcounter{model6} \setcounter{model6}{0}
\newcommand{\modelone}[1]{\noindent%
	\refstepcounter{model1}\text{(M1.\arabic{model1})}\\%
}
\newcommand{\modeltwo}[1]{\noindent%
	\refstepcounter{model2}\text{(M2.\arabic{model2})}\\%
}
\newcommand{\modelthree}[1]{\noindent%
	\refstepcounter{model3}\text{(M3.\arabic{model3})}\\%
}


\section{Introduction and motivation} \label{sec:introduction}
    Driven by the increase in the average global temperature due to the large concentration of Greenhouse Gases (GHGs) in the atmosphere and their direct impact on worsening the effects of climate change, several countries have committed to installing a large amount of low-emission renewable energy sources (RES). For example, the European Union pledges to install 450~GW of offshore wind and have solar PV as the backbone of its energy system by 2050~\cite{EU_1}, while reducing its CO$_{2}$ emission by 55~\% compared to 1990 and reaching carbon-neutrality by 2050~\cite{EU_2}. Worldwide, 11-15~TW of solar PV and 7-10~TW of wind are expected to be installed by 2050~\cite{IRENA}, when their cumulative installed capacity was 2~TW~\cite{GSC} and roughly 1~TW in 2024~\cite{WWEA}, respectively.
    
\begin{figure}[h!]
        \centering
        \includegraphics[width=0.65\linewidth]{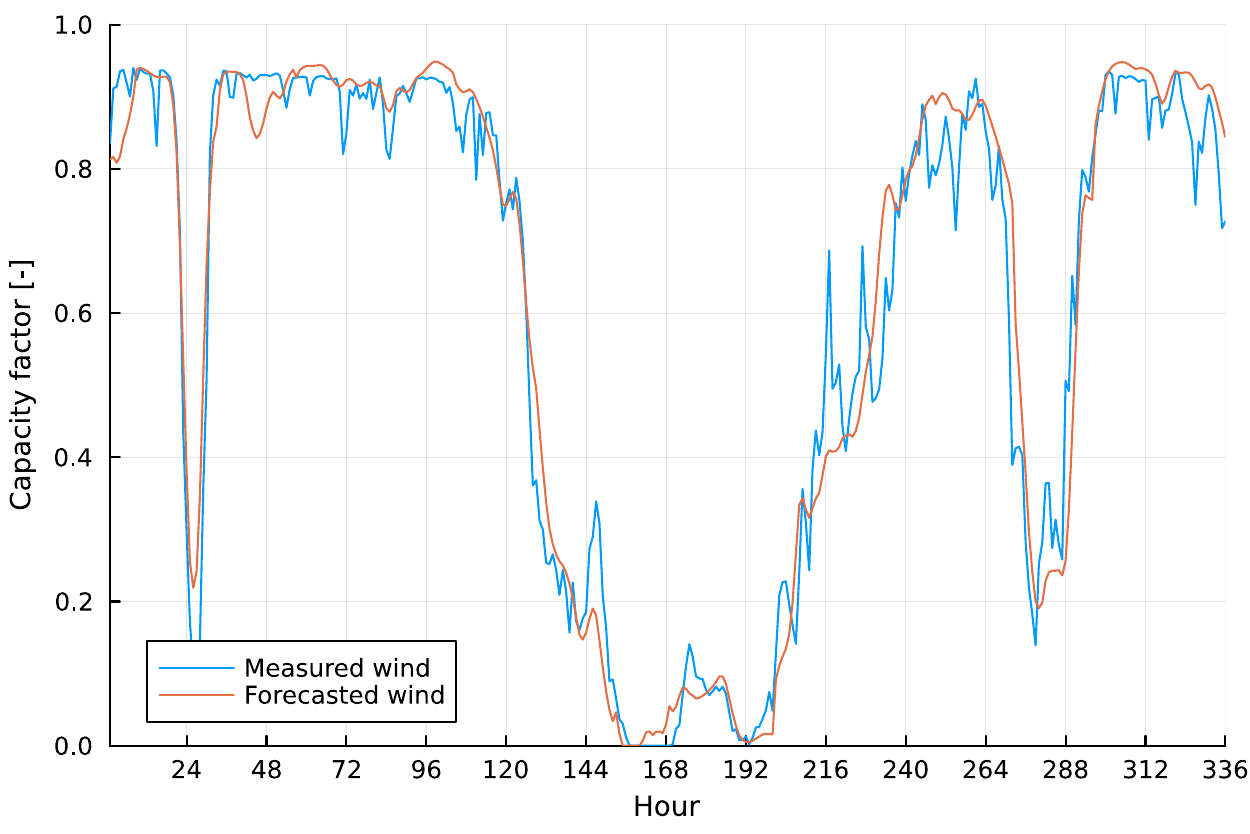}
        \caption{Example of a varying offshore wind capacity factor, corresponding to the measured wind power (blue) and day-ahead forecasted wind power (orange)  for the Belgian offshore wind farms, which have a total monitored capacity of 2.261~GW~\cite{Elia_2024_OFW}. D-1 forecasts for the next day are provided at 11am. }
        \label{fig:RES_variability}
    \end{figure}

    These RES provide power with considerably fewer emissions compared to conventional generators and are therefore excellent candidates for achieving a net-zero society in the future. However, they have a fluctuating nature directly related to the availability of resources in the areas where they are installed. This fluctuating nature results in rapidly changing power flows that cannot yet be perfectly forecasted, potentially causing sudden and significant variations in power generation within the power grid. An example of RES' variability is shown for a two-week-long period in Figure~\ref{fig:RES_variability}, where the offshore wind capacity factor values are shown. These values will be used later in the paper.
    
    In addition, power grids have historically been expanded to maximise the power transfer capacity between the demand centres and bulk conventional power plants, with recurrent and \textit{expected} power flows in the grid. Distributed large RES capacities cause variations in these recurrent power flows and potentially cause congestion in grids that were built without considering their distributed and fluctuating nature. As a result of these facts, the power grid is considered a bottleneck for the deployment of RES~\cite{IEA_2023} and can lead to massive RES curtailment if it is not expanded efficiently~\cite{JRC_2024}. In this regard, congestion costs have reached record-high numbers in recent years both in Europe~\cite{ACER_congestion} and the US~\cite{Congestion_US}, with Germany paying 2.774 bn€ in 2024 alone~\cite{DE_COSTS,SMARD}, and all the US Independent System Operators combined ranging from 13.5 to 8 bn\$ in 2022 and 2023, respectively~\cite{Congestion_US}.    
    In addition, the global electricity demand is rapidly increasing~\cite{IEA_2025}, and further delaying urgent grid reinforcements would result in additional losses for the socioeconomic welfare~\cite{IEA_2023,JRC_2024}, and more GHGs emissions. Despite this urgent need for investments in the power grid, transmission projects are often delayed because of technical (supply chain issues, lack of workforce, lack of manufacturing capacity, bottlenecks with critical elements, etc.~\cite{IEA_2025_constraints}) and non-technical (public opposition~\cite{NEUKIRCH_Grid_acceptance}) constraints. 
    
\begin{figure}[h!]
        \centering
        \includegraphics[width=0.7\linewidth]{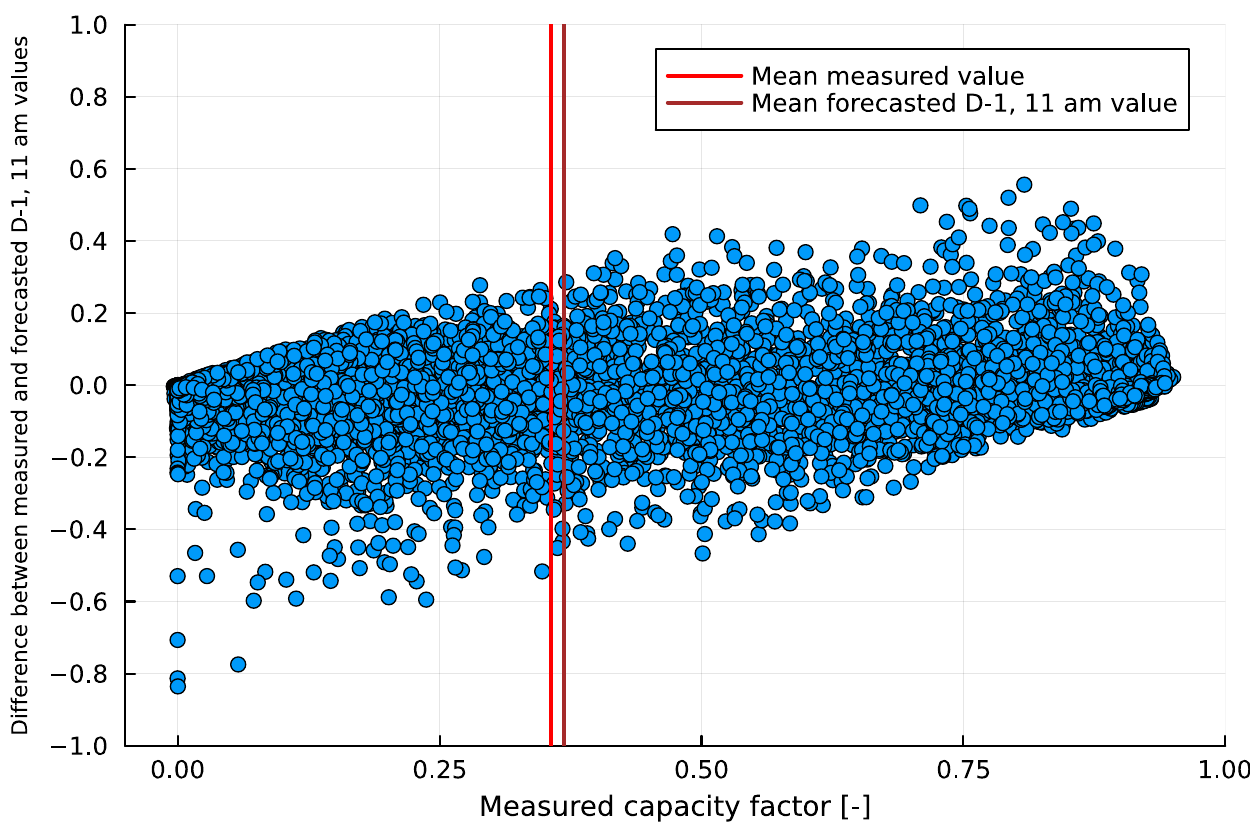}
        \caption{Difference between the measured and forecasted D-1, 11 am capacity factor values for all the quarterly measured capacity factor in 2024 for the Belgian offshore wind farms, monitored capacity of 2.261~GW~\cite{Elia_2024_OFW}.}
        \label{fig:Elia_2024_OFW}
    \end{figure}

    Based on these issues, congestion management actions such as power flow control are usually performed in Europe through phase-shifting transformers and high-voltage direct current (HVDC) links. Even though these actions are well-established, they are almost entirely limited to relieving cross-border congestion, while other means of non-costly remedial actions are needed to also deal with internal grid congestion. For this reason, several initiatives both in Europe~\cite{nrao} and the US~\cite{EPRI} propose to utilize the existing grid more efficiently by, among others, dynamically adjusting the grid topology to resolve congestion and maximize the power transfer capacity. The general idea is to use the power grid more efficiently without the need for expensive investments in it, while waiting for or fully replacing grid expansion projects. Nevertheless, these initiatives are currently not employed for congestion management in a dynamic manner, and the available specific, predefined topological actions rely mainly on expert-based knowledge from the past. Particularly, the Core Capacity Calculation Process used by ENTSO-E~\cite{ENTSO_E_CCR} focuses on the use of non-costly remedial action optimization to maximize the available capacity for cross-zonal exchanges between each of the European electricity market zones, starting the optimization process from two days-ahead (D-2) until one day-ahead (D-1), before the electricity market is cleared. Considering the large RES penetration expected in the future, it is mandatory to include the inherent uncertainty in the RES forecast in the grid topology optimization process. For example, as shown in Fig.~\ref{fig:Elia_2024_OFW} plotted from wind measurements from the OpenDataElia platform from the Belgian Transmission System Operator (TSO) Elia~\cite{Elia_2024_OFW}, the difference in capacity factor between the measured value and the D-1 forecasted value at 11 am for the year 2024 can be considerable for a wide range of measured capacity factors. 
    
    With an increased installed capacity, the difference between the measured and forecasted values might create imbalances in the grid, especially if the grid topology is optimized only considering the D-1 forecasted value for a large area. Therefore, as it will be shown later in the paper, considering the RES uncertainty in the grid optimization process helps to avoid using sub-optimal topologies, while achieving economical benefits compared to the original topology. Consequently, this work extends the deterministic steady-state grid topology optimization model for both AC and hybrid AC/DC grids in~\cite{Bastianel} by proposing a multistep \& stochastic model for a period of $\timesteps$ timesteps, taking into account RES' uncertainty through a scenario-based uncertainty representation through $\scenarios$ scenarios.

After this introduction and motivation section, the paper is organized as follows. Section~\ref{sec:literature_review} relates this work to relevant literature in the field of grid topology optimization, while Section~\ref{sec:contributions} lists the contributions of this paper.  Section~\ref{sec:methodology} presents both the proposed multistep \& stochastic grid topology optimization models, and how RES uncertainty is quantified in the paper. The input data (test cases and time series) to test the proposed methodology are described in Section~\ref{sec:input_data}, while Section~\ref{sec:results} presents the results of the proposed method based on the previously defined input data. Finally, Section~\ref{sec:conclusion} includes the conclusion and mentions ideas for future work.

\footnotetext[1]{https://github.com/Electa-Git/PowerModelsTopologicalActions.jl}
\footnotetext[2]{https://github.com/Electa-Git/StochasticPowerModelsTopologicalActions.jl}
\section{Related work} \label{sec:literature_review}
This Section deals with the two main contributions of the paper, i.e., formulating a multistep steady-state grid topology optimization model for AC and hybrid AC/DC grids and including RES uncertainty in the grid topology optimization model. 

The model formulated in this paper extends the single-step grid topology optimization model for hybrid AC/DC grids introduced in~\cite{Bastianel}, where optimal transmission switching (OTS) and busbar splitting (BS) can be performed on both AC and (DC) busbars. Previous work in the literature is mainly based on the linearized ``DC'' formulation of the OTS~\cite{Fisher2008,Hedman2009}, BS~\cite{Morsy2022,Hinneck2022,MORSY_2025}, and combined OTS/BS problems~\cite{Heidarifar2016}, which often fail to produce ``AC'' feasible solutions in OPF and OTS/BS problems~\cite{Baker2021}. Among the mentioned works, only Morsy et al.~\cite{MORSY_2025} include hybrid AC/DC grids in their formulation. Given the expected growth of HVDC interconnections and extended multi-terminal hybrid AC/DC grids especially in Europe, modeling them is a priority to efficiently operate the future grid. Moreover, the model in~\cite{Bastianel} is based on the linear programming AC approximation (LPAC) formulation and provides ``AC'' feasible solutions for the optimal topologies computed with the proposed combined OTS/BS problems. Furthermore, the LPAC formulation led to computationally faster and more accurate results compared to relaxation techniques used in the literature, such as the second-order cone (SOC)~\cite{Heidarifar2021} and the quadratic convex (QC) relaxations~\cite{QC,OTS_QC}. 

All the references mentioned above provide a formulation where the optimal transmission topology is determined for a single timestep. However, in real life, switching actions in the grid need to be scheduled carefully, and carried out by control room operators. As such, the number of switching actions that can be performed over a day is often pre-determined and limited. For modeling this behavior, existing optimization models need to be extended for multiple timesteps (dispatch periods), e.g. hours, by introducing intertemporal constraints. Li et al.~\cite{Li2019IEEEAccess} incorporate grid topology optimization and dynamic line rating (DLR) in a network-constrained ``DC''-unit commitment (UC) model to quantify their influence on the next-day power system scheduling, therefore optimizing the grid operations day-ahead for multiple timesteps. Numan et al.~\cite{Numan_2020} combine OTS and DLR with an optimal investment plan for building new transmission lines in a co-optimized network expansion planning model, but for a given single time horizon. Lai and Teh~\cite{LAI_2022} combine grid topology optimization with DLR, battery storage, and large RES penetration. The effect of RES variability on grid topology optimization is investigated in the context of unit commitment (UC) problems, such as in~\cite{Li2019IEEEAccess}. For example, Kawasaki~\cite{KAWASAKI2022} studies the flexibility offered by RES in a multi-step UC model with the grid topology that can be optimized. The same authors~\cite{KAWASAKI2024} use their model to reduce the short-circuit capacity in the system. A stochastic UC  problem with day-ahead wind uncertainty is tackled by Ruiz et al. in~\cite{RUIZ_2009} and by Papavasiliou et al. in~\cite{Papavasiliou_2013} for a transmission-constrained network with high wind penetration. Although there exist several seminal papers~\cite{Takriti_1996,Zheng_2015} discussing the application of stochastic models to the UC problem and related mathematical decompositions~\cite{Carpentier_1996}, or with parallel ways of treating the uncertainty, e.g, with chance-constrained programming~\cite{Ozturk_2004}, they do not include optimal transmission switching or bus bar splitting actions. In this context, a chance-constrained linearized ``DC'' model is used to add uncertainty from RES to the OTS in~\cite{Qiu_2015}, while Han et al. add the uncertainty related to RES and N-K contingencies to an OTS piecewise linear model via a multi-stage model including stochastic and distributionally robust optimization~\cite{Han_2023}.

The following Table~\ref{table:ref} includes an overview of all the grid topology optimization-related literature mentioned in this Section. As indicated in the last row, this paper represents the first attempt to include RES uncertainty in a multistep (24-hour-long) grid topology optimization model combining both OTS and BS. Even without RES uncertainty, there is limited literature dealing with multistep (\& stochastic) steady-state grid topology optimization problems, let alone for hybrid AC/DC grids. 
\begin{table}[h!]
\caption{Overview of multistep \& stochastic topological actions literature. ``Grid topology optimization'' includes both optimal transmission switching (OTS) and busbar splitting (BS). The different levels of complexity are mixed-integer linear programming (MILP), mixed-integer quadratic convex programming (MIQCP), and mixed-integer second-order cone programming (MISOCP). * indicates that even though the formulation for~\cite{Bastianel} and for this paper is a piecewise linear approximation, the model is implemented with quadratic constraints for computational efficiency.}
\centering
{\fontsize{7pt}{11pt}\selectfont
\begin{tabular}{c|cc|c|c|c|c|cc}
\hline
  & \multicolumn{2}{c|}{Grid topology} & & &  & &   \\
Reference & \multicolumn{2}{c|}{optimization} & OPF formulation & Complexity & Multistep & Stochastic &  \multicolumn{2}{c}{Grid type} \\
  & OTS & BS & & & & & AC & AC/DC     \\
\hline
\cite{ENTSO_E_CCR}        & \textbf{\ding{55}} & \textbf{\ding{51}} & Linear approximation (DC) & MILP & \textbf{\ding{51}} & \textbf{\ding{55}} & \textbf{\ding{51}} & \textbf{\ding{55}}\\
\cite{Bastianel}          & \textbf{\ding{51}} & \textbf{\ding{51}} & Piecewise linear approximation (LPAC) & MIQCP* & \textbf{\ding{55}} & \textbf{\ding{55}} & \textbf{\ding{51}} & \textbf{\ding{51}}\\ 
\cite{Fisher2008}         & \textbf{\ding{51}} & \textbf{\ding{55}} & Linear approximation (DC) & MILP & \textbf{\ding{55}} & \textbf{\ding{55}} & \textbf{\ding{51}} & \textbf{\ding{55}}\\ 
\cite{Hedman2009}         &\textbf{\ding{51}}  & \textbf{\ding{55}} & Linear approximation (DC) & MILP & \textbf{\ding{55}} & \textbf{\ding{55}} & \textbf{\ding{51}} & \textbf{\ding{55}}\\ 
\cite{Morsy2022}          & \textbf{\ding{55}} & \textbf{\ding{51}} & Linear approximation (DC) & MILP & \textbf{\ding{55}} & \textbf{\ding{55}} & \textbf{\ding{51}} & \textbf{\ding{55}}\\  
\cite{Hinneck2022}        & \textbf{\ding{55}} & \textbf{\ding{51}} & Linear approximation (DC) & MILP & \textbf{\ding{55}} & \textbf{\ding{55}} & \textbf{\ding{51}} & \textbf{\ding{55}}\\ 
\cite{MORSY_2025}         & \textbf{\ding{51}} & \textbf{\ding{51}} & Linear approximation (DC) & MILP & \textbf{\ding{55}} & \textbf{\ding{55}} & \textbf{\ding{55}} & \textbf{\ding{51}} \\ 
\cite{Heidarifar2016}     & \textbf{\ding{51}} & \textbf{\ding{51}} & Linear approximation (DC) & MILP  & \textbf{\ding{55}} & \textbf{\ding{55}} & \textbf{\ding{51}} & \textbf{\ding{55}} \\ 
\cite{Heidarifar2021}     & \textbf{\ding{51}} & \textbf{\ding{51}} & Second order cone relaxation (SOC) & MISOCP & \textbf{\ding{55}} & \textbf{\ding{55}} & \textbf{\ding{51}} & \textbf{\ding{55}} \\  
\cite{OTS_QC}                 & \textbf{\ding{51}} & \textbf{\ding{55}} & Quadratic convex relaxation (QC) & MIQCP & \textbf{\ding{55}} & \textbf{\ding{55}} & \textbf{\ding{51}} & \textbf{\ding{55}} \\ 
\cite{Li2019IEEEAccess}   & \textbf{\ding{55}} & \textbf{\ding{51}} & Linear approximation (DC) & MILP & \textbf{\ding{51}} & \textbf{\ding{55}} & \textbf{\ding{51}} & \textbf{\ding{55}} \\ 
\cite{Numan_2020}         & \textbf{\ding{51}} & \textbf{\ding{55}} & Linear approximation (DC) & MILP  & \textbf{\ding{55}} & \textbf{\ding{51}} & \textbf{\ding{51}} & \textbf{\ding{55}} \\ 
\cite{LAI_2022}           & \textbf{\ding{51}} & \textbf{\ding{55}} & Linear approximation (DC) & MILP  & \textbf{\ding{51}} & \textbf{\ding{55}} & \textbf{\ding{51}} & \textbf{\ding{55}} \\ 
\cite{KAWASAKI2022}       & \textbf{\ding{51}} & \textbf{\ding{55}} & Linear approximation (DC) & MILP & \textbf{\ding{51}} & \textbf{\ding{55}} & \textbf{\ding{51}} & \textbf{\ding{55}} \\ 
\cite{KAWASAKI2024}       & \textbf{\ding{51}} & \textbf{\ding{55}} & Linear approximation (DC) & MILP & \textbf{\ding{51}} & \textbf{\ding{55}} & \textbf{\ding{51}} & \textbf{\ding{55}} \\ 
\cite{Qiu_2015}           & \textbf{\ding{51}} & \textbf{\ding{55}} & Linear approximation (DC) & MILP & \textbf{\ding{55}} & \textbf{\ding{51}} & \textbf{\ding{51}} & \textbf{\ding{55}} \\ 
\cite{Han_2023}           & \textbf{\ding{51}} & \textbf{\ding{55}} & Piecewise linear approximation (LPAC) & MILP & \textbf{\ding{55}} & \textbf{\ding{51}} & \textbf{\ding{51}} & \textbf{\ding{55}} \\  
\hline
This paper               & \textbf{\ding{51}} & \textbf{\ding{51}} & Piecewise linear approximation (LPAC) & MIQCP* & \textbf{\ding{51}} & \textbf{\ding{51}} & \textbf{\ding{51}} & \textbf{\ding{51}} \\ 
\end{tabular}
}
\label{table:ref}
\end{table}

\section{Contributions} \label{sec:contributions}
Based on the motivation and the literature mentioned in the previous sections, this paper extends the state-of-the-art steady-state grid topology optimization model for AC grids and hybrid AC/DC grids introduced in~\cite{Bastianel} by:
\begin{itemize}
    \item Adding a multistep \& stochastic model to optimize the topology for a period of $\timesteps$ hours and $\scenarios$ scenarios for each hour. The model allows for:
    \begin{itemize}
        \item Optimizing the grid topology hourly, taking into account RES' uncertainty through the $\scenarios$ scenarios per hour.
        \item Selecting one single optimal topology valid over a given time period, taking into account RES' uncertainty through the $W$ scenarios per hour.
        \item Performing up to $S^{max}$ switching actions $s_{t}$ in a given time period of 24 hours, with $SA^{max}$ selected by the user, taking into account RES' uncertainty through the $W$ scenarios per hour.
    \end{itemize}
    Note that the model can also optimize only one of the two dimensions, i.e. a single timestep model ($T$ = 1) with $W$ scenarios of a multistep ($T >$ 1) model with no stochasticity ($W$ = 1).
    \item Adding a methodology based on K-means clustering to quantify RES uncertainty through a scenario-based stochastic optimization using real forecasted offshore wind data used by TSOs~\cite{Elia_2024_OFW}.
    \item Testing the results of the 
    Piecewise Linear (LPAC)~\cite{LPAC} grid topology optimization model implemented as a Mixed-Integer Quadratic Convex Problem (MIQCP) in~\cite{Bastianel} for feasibility with the nonlinear, non-convex ``AC-OPF'' formulation~\cite{PowerModels2018,ergun_optimal_2019}, and running a redispatch model if the D RES capacity factor differs from the D-1 forecast.
\end{itemize}

Note that all the models are made publicly available in the PowerModelsTopologicalActions.jl\footnotemark[1] and StochasticPowerModelsTopologicalActions.jl\footnotemark[2] Julia packages.

\section{Methodology} \label{sec:methodology}
As introduced in Section~\ref{sec:introduction}, the large integration of RES in the power system creates novel power flows compared to the \textit{expected} ones that characterized conventional generation-dominated grids. As those traditional power systems were quite static, grid congestion often occurred on the same lines, and there was therefore one optimal dominant topology that was adjusted mainly in case of contingencies, following actions based on expert-based knowledge from the TSO.
This paper aims to investigate whether the \textit{less expected} power flows caused by the RES variability (and inherent RES forecast uncertainty) increase the value for topology optimization, as large-scale changes in power flow patterns might occur across hours/days, and new topologies might be considered to solve grid congestion and minimize the total generation costs.

\subsection{Grid topology optimization model}
This section describes the optimization models developed within this work. After defining the sets in Section~\ref{subsec:sets}, the LPAC-BS model from~\cite{Bastianel} is introduced in Section~\ref{subsec:multistep_stochastic}. Section~\ref{subsec:multistep_stochastic} describes the multistep \& stochastic model, while Section~\ref{subsec:workflow} includes the workflow of the proposed grid topology optimization model.

\subsubsection{Sets}~\label{subsec:sets}
As this work extends the models built in~\cite{PowerModels2018},~\cite{ergun_optimal_2019} and~\cite{Bastianel}, it uses the same terminology for network components, topologies, connectivities, and nomenclature to distinguish each network component unambiguously. By using `reverse' topologies for AC and DC branches, the same branches described before can represent flows in the opposite direction. Note that both radial and meshed topologies (both on the AC and DC sides) are supported by the proposed model.
\begin{table}[!ht]
	\renewcommand{\arraystretch}{1.1}
        {\fontsize{9pt}{10pt}\selectfont
	\centering
	\begin{tabular}{m{37em} l}
		\hline
		Network components &\\
		\hline
            $\acnodes$          & Set of AC nodes             \\           
            $ \acbranches $     & Set of AC branches  \\          
            $ \acswitches $     & Set of AC switches  \\          
            $ \dcnodes $        & Set of DC nodes        \\          
            $ \dcbranches $     & Set of DC branches  \\          
            $ \dcswitches $     & Set of DC switches  \\          
            $ \acdcconverters $ & Set of AC/DC converters  \\
            $ \generators $     & Set of AC generators   \\
            $ \loads $          & Set of AC loads   \\
            $ \acnodesnew $     & Set of AC nodes for BS \\
            $ \acZIL $          & Set of AC busbar couplers  \\
            $ \dcnodesnew $     & Set of DC nodes for BS  \\
            $ \dcZIL $          & Set of DC busbar couplers  \\
 		\hline
		Topologies and connectivities &\\
		\hline
            $\actopology$         & AC topologies             \\
            $\actopologyrev$      & Reverse AC topologies     \\ 
            $ \dctopology $       & DC topologies             \\
            $ \dctopologyrev$     & Reverse DC topologies     \\
            $ \acswitchtopology $ & AC switch topologies      \\
            $ \dcswitchtopology $ & DC switch topologies      \\
            $ \acZILtopology $    & AC zero impedance lines   \\    
            $ \dcZILtopology $    & DC zero impedance lines   \\   
            $ \convertertopology$ & AC/DC converter topologies\\
            $ \acloadconn $       & AC load connectivity      \\
            $ \genconn $          & AC generator connectivity \\
            \hline
		Connectivity of each network component &\\
            \hline
            $lij~\in~\actopology~\subseteq~\acbranches~\times~\acnodes~\times~\acnodes$ & AC branches to AC nodes \\
            $\upsilon ii'~\in~\acZILtopology \subseteq~\acswitches~\times~\acnodes~\times~\acnodesnew$ & AC busbar couplers \\
            $\upsilon mi~\in~\acswitchtopology\subseteq~\acswitches~\times~\acnodes~\times~\acnodes$ & AC switches \\
            $def~\in~\dctopology~\subseteq~\dcbranches~\times~\dcnodes~\times~\dcnodes$ & DC branches \\ 
            $\xi ee'~\in~\dcZILtopology~\subseteq~\dcswitches~\times~\dcnodes~\times~\dcnodesnew$ & DC busbar couplers \\
            $\xi re~\in~\dcswitchtopology~\subseteq~\dcswitches~\times~\dcnodes~\times~\dcnodes$ & DC switches \\
            $lji~\in~\actopologyrev~\subseteq~\acbranches~\times~\acnodes~\times~\acnodes$ & Reverse AC branches \\
            $dfe~\in~\dctopologyrev~\subseteq~\dcbranches~\times~\dcnodes~\times~\dcnodes$ & Reserve DC branches \\

            $mi~\in~\acloadconn$ & Loads \\
            $gi~\in~\genconn$ & Generators \\
            $cie~\in~\convertertopology$ & AC/DC Converters \\
            \hline
            Time- and uncertainty-related components &\\
            \hline
            $t~\in~\timesteps$ & Timesteps \\
            $w~\in~\scenarios$ & Scenarios for each timestep \\
            $\omega~\in~\Omega$ & Probability for each scenario $w$ \\
            \hline
	\end{tabular}
    }
\end{table}


\subsubsection{Multistep and stochastic model} \label{subsec:multistep_stochastic}
While the ``AC''-OPF model for hybrid AC/DC grids is well described in the literature~\cite{ergun_optimal_2019,Bastianel}, the ``LPAC'' formulation of the combined OTS and BS model (LPAC-BS) from~\cite{Bastianel} is introduced in this work for clarity. To keep the text concise, Model 1 presented thereafter includes the model from~\cite{Bastianel} with $T$ timesteps and $\scenarios$ scenarios for each timestep $t$. The original model can be obtained by removing the multi-step and stochastic dimensions from Model 1, i.e., removing the $T$ timesteps and $\scenarios$ scenarios indices from (M1.1)-(M1.28). This LPAC-BS model is based on the LPAC-OPF implementation in~\cite{PowerModels2018,ergun_optimal_2019}.

\begin{table}[!ht]
	\renewcommand{\arraystretch}{1.0}
        {\fontsize{9pt}{10pt}\selectfont
	\centering
	\label{tb:Model1}
	\begin{tabular}{m{29em} r}
		\hline
		Model 1: Stochastic multistep LPAC-BS model for hybrid AC/DC grids & (* refers to the DC part of the hybrid AC/DC grid)\\
		\hline
		\textbf{Minimize:}\\
		 $\sum_{t \in T}(\sum_{w \in W}\omega(\sum_{k \in G} c_{2k}\cdot {P_{k,w,t}^{g}}^{2} + c_{1k}\cdot P_{k,w,t}^{g} + c_{0k}) + \sum_{ZILii' \in \acZILtopology} c_{sw}\cdot z^{sw,ac}_{ZILii',t} + \sum_{ZILee' \in \acZILtopology} c_{sw}\cdot z^{sw,dc}_{ZILee',t})$ & \modelone \\
		\textbf{AC bus:}\\		
          $\theta_{r,t,w} = 0 $ & \modelone \\
          $(\underline{U}^{m}_{i} - 1) \leq \phi_{i,w,t} \leq (\overline{U}^{m}_{i} - 1)$ & $\forall i \in \acnodes, \forall t \in T, \forall w \in \scenarios $  \quad \modelone\\  
          $ \underline{\theta}_{i} \leq \theta_{i,w,t} \leq \overline{\theta}_{i}$ & $\forall i \in \acnodes, \forall t \in T, \forall w \in \scenarios$  \quad \modelone \\
            $\sum_{\substack{k \in \generators_i}} S^g_{k,w,t} + \sum_{\substack{l \in \loads_{i}}} S^m_{l,w,t} - (\sum_{\substack{c \in \acdcconverters_{i}}} S^{cv,ac}_{c,w,t})^{*} - Y^s_{i,w,t}(1+2\phi_{i,w,t}) + \sum_{\substack{\upsilon mi \in (\acswitchtopology_{i} \cup \acZILtopology_{i})}} {S^{sw,ac}_{\upsilon mi,w,t}} = \sum_{\substack{lij\in \actopology}} S^{ac}_{lij,w,t}$ & $ \forall i \in \acnodes, \forall t \in T, \forall w \in \scenarios$ \quad \modelone \\    
          \textbf{Generator}&\\
          $\underline{S}^{g}_{k} \leq S^g_{k,w,t} \leq  \overline{S}^{g}_{k} $&$ \forall k \in \generators, \forall t \in T, \forall w \in \scenarios $ \quad \modelone\\
          \textbf{AC branch}&\\
            $P_{lij,w,t} = (g_{s,ij} + g_{ij})(1 + 2\cdot\phi_{i,w,t}) -g_{ij} (cs_{ij,wt} + \phi_{i,w,t} + \phi_{j,w,t}) - b_{ij}(\theta_{i,w,t}-\theta_{j,w,t}) $&$ \forall lij\in \actopology \cup \actopologyrev, \forall t \in T, \forall w \in \scenarios $  \quad \modelone \\
           $Q_{lij,w,t} = (b_{s,ij} + b_{ij})(1 + 2\cdot\phi_{i,w,t}) -b_{ij} (cs_{ij,wt} + \phi_{i,w,t} + \phi_{j,w,t}) - g_{ij}(\theta_{i,w,t}-\theta_{j,w,t}) $&$ \forall lij \in \actopology \cup \actopologyrev, \forall t \in T, \forall w \in \scenarios $  \quad \modelone \\
        $|P_{lij,w,t}| \leq  \overline{P}_{lij} $&$ \forall lij \in \actopology \cup \actopologyrev, \forall t \in T, \forall w \in \scenarios $ \quad \modelone\\
        $|Q_{lij,w,t}| \leq  \overline{Q}_{lij} $&$ \forall lij \in \actopology \cup \actopologyrev, \forall t \in T, \forall w \in \scenarios $ \quad \modelone\\
        $cs_{ij,wt} \leq 1 - \frac{(1 - cos(\overline{\Delta\theta_{ij}})}{(\overline{\Delta\theta_{ij}})^{2}}(\theta_{i,w,t}-\theta_{j,w,t})^{2} $&$ \forall lij \in \actopology \cup \actopologyrev, \forall t \in T, \forall w \in \scenarios$  \quad \modelone\\
        \textbf{AC/DC converter*}&\\
        $\underline{S}^{cv,ac}_c \leq S^{cv,ac}_{c,w,t} \leq  \overline{S}^{cv,ac}_c $&$ \forall c \in \acdcconverters, \forall t \in T, \forall w \in \scenarios$   \quad \modelone\\ 
        $ P^{cv,ac}_{c,w,t} + P^{cv, dc}_{c,w,t} = a + b |I^{cv,dc}_{c,w,t}| $&$ \forall c \in \acdcconverters, \forall t \in T, \forall w \in \scenarios$   \quad \modelone\\
        $c\cdot P^{cv,ac}_{c,w,t} + d \cdot Q^{cv,ac}_{c,w,t} \leq (S^{cv,ac}_{c,w,t})^2 $&$ \forall c \in \acdcconverters, \forall t \in T, \forall w \in \scenarios$ \quad \modelone \\
        \textbf{DC bus*} & \\
        $\underline{U}^{dc}_{e} \leq U^{dc}_{e,w,t} \leq \overline{U}^{dc}_{e} $&$ \forall e \in \dcnodes, \forall t \in T, \forall w \in \scenarios $  \quad \modelone\\  
        $\sum_{\substack{k \in \dcgenerators_{e,w,t}}} P^{g,dc}_{k,w,t} + \sum_{\substack{c \in \acdcconverters_e}} P^{cv, dc}_{c,w,t} - \sum_{\substack{l \in \dcloads}} {P^{m,dc}_{l,w,t}} - Y^s_{e,w,t}(1+2\phi_{e,w,t}) + \sum_{\substack{\xi re \in (\dcswitchtopology_{e} \cup \dcZILtopology_{e})}} {P^{sw,dc}_{\xi re,w,t}} = \sum_{\substack{def} \in \dctopology} P^{dc}_{def,w,t} $&$ \forall e \in \dcnodes, \forall t \in T, \forall w \in \scenarios$  \quad \modelone \\ 
        \textbf{DC branch*}&\\
        $P^{dc}_{def,w,t} =  p^{dc}_{def}  Y_{def} ( \phi^{dc}_{e,w,t} - \phi^{dc}_{f,w,t}) $&$  \forall def \in \dctopology \cup \dctopologyrev, \forall t \in T, \forall w \in \scenarios$    \quad \modelone \\
        $\underline{P}^{dc}_{def} \leq P^{dc}_{def,w,t} \leq \overline{P}^{dc}_{def} $&$ \forall def \in \dctopology \cup \dctopologyrev, \forall t \in T, \forall w \in \scenarios$    \quad \modelone\\
        \textbf{AC switches}&\\
        $ - (1 - z^{sw,ac}_{\upsilon mi,t}) \cdot M_{\theta} \leq \theta_{m,w,t} - \theta_{i,w,t} \leq (1 - z^{sw,ac}_{\upsilon mi,t}) \cdot M_{\theta} $&$ \forall \upsilon mi \in \acswitchtopology \cup \acZILtopology, \forall t \in T, \forall w \in \scenarios$  \quad \modelone\\
        $ - (1 - z^{sw,ac}_{\upsilon mi,t}) \cdot M_{m} \leq \phi_{m,w,t} - \phi_{i,w,t} \leq (1 - z^{sw,ac}_{\upsilon mi,t}) \cdot M_{m}  $&$ \forall \upsilon mi \in \acswitchtopology \cup \acZILtopology, \forall t \in T, \forall w \in \scenarios $  \quad \modelone \\
        $z^{sw,ac}_{\upsilon  mi,t} \cdot \underline{P}^{sw,ac}_{\upsilon  mi} \leq P^{sw,ac}_{\upsilon  mi,w,t} \leq z^{sw,ac}_{\upsilon  mi,t} \cdot \overline{P}^{sw,ac}_{\upsilon mi} $&$ \forall \upsilon mi \in \acswitchtopology \cup \acZILtopology, \forall t \in T, \forall w \in \scenarios $  \quad \modelone \\
        $z^{sw,ac}_{\upsilon mi,t} \cdot \underline{Q}^{sw,ac}_{\upsilon mi} \leq Q^{sw,ac}_{\upsilon mi,w,t} \leq z^{sw,ac}_{\upsilon mi,t} \cdot \overline{Q}^{sw,ac}_{\upsilon mi} $&$  \forall \upsilon mi \in \acswitchtopology \cup \acZILtopology, \forall t \in T, \forall w \in \scenarios $  \quad \modelone \\
        $z^{sw,ac}_{\upsilon mi,t} + z^{sw,ac}_{\kappa mi',t} \leq 1 $&$ \forall (\upsilon mi, \kappa mi') \in \acswitchtopology, \forall t \in $  \quad \modelone  \\
        $z^{sw,ac}_{\kappa mi',t} \leq (1 - z^{sw,ac}_{ZILii',t}) $&$ \forall (\upsilon mi, ZILii') \in \acswitchtopology \cup \acZILtopology, \forall t \in T $ \quad \modelone  \\
        \textbf{DC switches*}&\\
        $ - (1 - z^{sw,dc}_{\xi re,t}) \cdot M_{dc} \leq \phi^{dc}_{r,w,t} - \phi^{dc}_{e,w,t} \leq (1 - z^{sw,dc}_{\xi re,t}) \cdot M_{dc}$&$  \forall \xi re \in \dcswitchtopology \cup \dcZILtopology, \forall t \in T, \forall w \in \scenarios $  \quad \modelone \\
        $z^{sw,dc}_{\xi re,t} \cdot \underline{P}^{sw,dc}_{re} \leq P^{sw,dc}_{re,w,t} \leq z^{sw,dc}_{\xi re,t} \cdot \overline{P}^{sw,dc}_{\xi re} $&$ \forall \xi re \in \dcswitchtopology \cup \dcZILtopology, \forall t \in T, \forall w \in \scenarios$   \quad \modelone \\
        $z^{sw,dc}_{\xi re,t} + z^{sw,dc}_{\iota re',t} \leq 1 $&$ \forall \xi re,\iota re' \in \dcswitchtopology, \forall t \in T$  \quad \modelone \\
        $z^{sw,dc}_{\iota re',t} \leq (1 - z^{sw,dc}_{ZILee',t}) $&$ \forall (\xi re, ZILee') \in \dcswitchtopology \cup \dcZILtopology, \forall t \in T $ \quad \modelone \\
        \hline
	\end{tabular}
    }
\end{table}

The objective function in (M1.1) consists of minimizing the total costs of the $ \generators $ set of generators,
where $P_{k}^{g}$ is the active power injection. The probability $\omega$ of each scenario happening is added to the objective function, too. (M1.2) sets the voltage angle of the reference bus to zero while (M1.3) and (M1.4) limit respectively the voltage magnitude deviation and angles of the other buses between a minimum [$(\underline{U}^{m}_{i} - 1)$,$\underline{\theta_{i}}$] and a maximum [$(1- \overline{U}^{m}_{i})$,$\overline{\theta_{i}}$] value. Similarly, the generator setpoints are bounded by $[\underline{S}^{g}_{k},\overline{S}^{g}_{k}]$ in (M1.6). Moreover, the AC power balance is shown by (M1.5), where $S^{cv,ac}_{c,w,t}$
refers to injections from HVDC converters, and $S^m_{l,w,t}$ to the
nodal demand. The term $Y_{i,w,t}^{s} \cdot (1+2\phi_{i,w,t})$ refers to the power absorbed by shunt elements connected to the node $i$. They are considered negligible in the remainder of the paper. The active ($P_{lij,w,t}$) and reactive power ($Q_{lij,w,t}$) flows through the AC branches are expressed using (M1.7 - M1.9), following the equations defined for the LPAC formulation in~\cite{PowerModels2018,LPAC}. The active and reactive powers are constrained by (M1.9) and (M1.10), while 
the approximation $cs$ of $cos(\theta_{n,w,t} - \theta_{m,w,t})$ for the LPAC formulation~\cite{LPAC} is expressed in (M1.11). The apparent power through the converter AC side is limited between the bounds $\underline{S}^{cv,ac}_{c,w,t}$ and $\overline{S}^{cv,ac}_{c,w,t}$ by (M1.12).

On the DC side of the grid, the power balance for the DC nodes $e$ in $ \dcnodes $ is defined by (M1.16). As for the power balance in the AC side in (M1.5), $P^{g,dc}_{k,w,t}$ refers to DC generators, $P^{cv, dc}_{c,w,t}$ to the DC side of the AC/DC converters, $P^{d,dc}_{e,w,t}$ to the DC loads, $Y^s_{e,w,t} \cdot (1 + 2\phi_{e,w,t})$ to the shunt elements and $P^{dc}_{def,w,t}$ to the DC branches. The power flow through the DC branches is represented in (M1.17), where $p^{dc}_{def}$ is the number of DC poles\footnote{1 for monopolar HVDC and 2 for symmetrical monopoles and bipoles}. The power flow through the DC branches is bounded within $[\underline{P}^{dc}_{def},\overline{P}^{dc}_{def}]$ by (M1.18). Finally, the AC/DC converters' AC and DC sides are linked by (M1.13), expressing the losses of the converter dependent on the AC side converter current $I^{cv}_{c,w,t}$, and the parameters $a$ and $b$. The AC/DC converter apparent power $S^{cv,ac}_{c,w,t}$ on the DC side is related to the active and reactive power of the converter through the parameters $c$ and $d$, fully defined in~\cite{ergun_optimal_2019}. As the converter transformer and filters can be modeled as AC branches without loss of generality, they are not further described in this paper. The interested reader is referred to~\cite{ergun_optimal_2019} for a full description of the AC/DC converter model. 

The AC and DC switches ${sw_{\upsilon mi,t}^{ac}}$, ${sw_{\upsilon mi'}^{ac}}$ and ${sw_{ZILii',t}^{ac}}$, and $z^{sw,dc}_{\xi re,t}$, $z^{sw,dc}_{\xi re',t}$ and $z^{sw,dc}_{ZIL ee',t}$ are subject to (M1.19)-(M1.28) \cite{Bastianel}.
(M1.19),(M1.20) and (M1.26) are related to the voltage angles (only AC) and the voltage magnitudes of the two buses at the extremes of each AC ($sw^{ac}_{\upsilon mi,t}$) and DC ($z^{sw,dc}_{\xi re,t}$) switch. (M1.21),(M1.22) and (M1.27) limit the active and reactive powers of the AC and DC switches to its maximum $\overline{P}^{sw,ac}_{\upsilon mi,w,t}$, $\overline{Q}^{sw,ac}_{\upsilon mi,w,t}$, $\overline{P}^{sw,dc}_{\xi re,w,t}$ and minimum values $\underline{P}^{sw,ac}_{\upsilon mi,w,t}$, $\underline{Q}^{sw,ac}_{\upsilon mi,w,t}$, $\underline{P}^{sw,dc}_{\xi re,w,t}$. (M1.23) and (M1.25) are ``exclusivity" constraints including the switches connecting each grid element to the split busbar. They are either equality ($=$ 1) or inequality ($\leq$ 1) constraints depending on whether OTS is performed or not. If OTS and BS are both allowed in the same optimization problem, (M1.23) and (M1.27) are inequality constraints, and both switches are allowed to be open. If (M1.23) and (M1.27) are equality constraints, each grid element decoupled from the original busbar needs to be reconnected to one part of the split busbar, and one of the two switches must be closed. Note that constraints (M1.24) and (M1.28) impose that if the busbar coupler is closed, i.e. busbar splitting is not performed, the switches $sw^{ac}_{\upsilon mi,t}$ and/or $sw^{dc}_{\xi re,t}$ connecting the network element to the original busbar are always closed. While both AC and DC switches are present in Model 1, one can select to use either only AC or only DC switches, depending on the busbars to be split.

\begin{figure}[h!]
    \centering
    \includegraphics[width=0.9\linewidth]{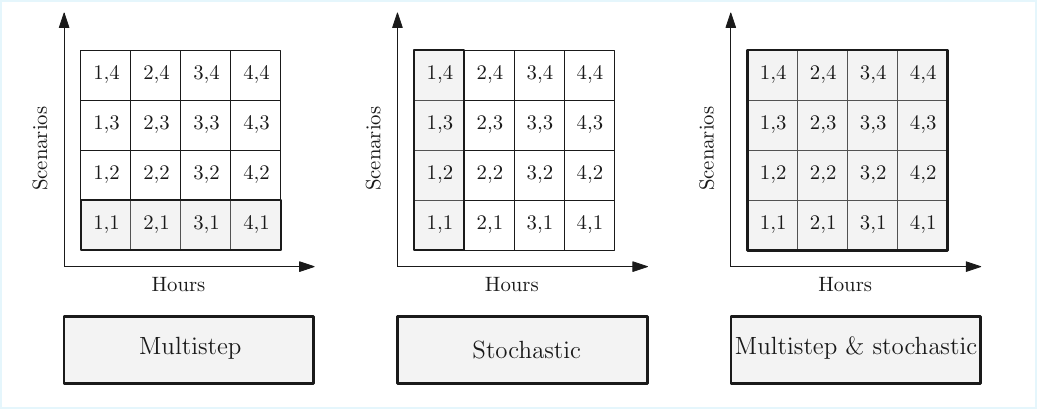}
    \caption{Rational behind the different formulations introduced in this paper. A multistep model (left) has $>$ 1 timesteps $t$ with 1 scenario $w$. The stochastic model (center) has 1 timestep $t$ and $W$ scenarios. The multistep \& stochastic model (right) has $T$ timesteps and $W$ scenarios for each timestep $t$.}
    \label{fig:illustration}
\end{figure}

Note that the binaries representing the status of each AC ($z^{sw,ac}_{\upsilon mi,t}$) and DC ($z^{sw,dc}_{\xi re,t}$) switch are only dependent on the timestep $t$, as they are consistent over each scenario $w$ in timestep $t$. One can study the topology optimization for only one of the two dimensions, i.e. a single-step model ($T$ = 1) with $W$ scenarios or a multistep ($T >$ 1) model with no stochasticity ($W$ = 1). In the first case, the $t$ index is removed from every variable in Model 2. In the latter case, the $w$ index is removed from each variable, and the probability $\omega$ in the timestep is 1. The possible different relations between hours and scenarios are summarised in Fig.~\ref{fig:illustration}.

Furthermore, Model 2 introduces additional constraints to the formulation introduced in Model 1:
\begin{itemize}
    \item (M2.1) fixes the status of each switch over the simulation length $\timesteps$. As a result, only one optimal topology is allowed in each simulation with $\timesteps$ timesteps. 
    \item (M2.2) to (M2.7) limit the number of possible switching actions ($s_{t}$,$s^{dc}_{t}$) over the simulation length $\timesteps$ to a value $SA^{max}$,$SA^{dc,max}$ for both AC ((M2.2)-(M2.4)) and DC ((M2.5)-(M2.7)) switches. Note that all the switches have 'slots' $s_{t}$ and $s^{dc}_{t}$ in which they can change their status together according to (M2.4) and (M2.7) for respectively AC and DC switches.
\end{itemize}

\begin{table}[h]
	\renewcommand{\arraystretch}{1.0}
    	{\fontsize{9pt}{10pt}\selectfont
	\centering
	\label{tb:Model1}
	\begin{tabular}{m{31em} r}
		\hline
		Model 2: Additional constraints for busbar splitting &\\
		\hline
        \textbf{One topology}\\
        $z^{sw,ac}_{\upsilon mi,t} = z^{sw,ac}_{\upsilon mi,t-1} $&$ \forall \upsilon mi \in \acswitchtopology, \forall t \in 2:T$ \quad \quad \modeltwo \\       
        \hline
        \textbf{Limiting the number of switching actions $s$ in a given time period} \\
        - AC Busbar Splitting:&\\
        $z^{sw,ac}_{\upsilon mi,t} - z^{sw,ac}_{\upsilon mi,t-1} \leq s_{t} $&$ \forall \upsilon mi \in \acswitchtopology \cup \acZILtopology, \forall t \in 2:T$ \quad \modeltwo  \\
        $z^{sw,ac}_{\upsilon mi,t-1} - z^{sw,ac}_{\upsilon mi,t} \leq s_{t}
        $&$ \forall \upsilon mi \in \acswitchtopology \cup \acZILtopology, \forall t \in 2:T $ \quad \modeltwo \\
        $\sum_{t=1}^{T} s_{t} \leq S^{max} $&$ \forall t \in T $  \quad \modeltwo \\
        - DC Busbar Splitting:&\\
        $z^{sw,dc}_{\xi re,t} - z^{sw,dc}_{\xi re,t-1} \leq s^{dc}_{t} $&$ \forall \xi re \in \dcswitchtopology \cup \dcZILtopology, \forall t \in 2:T$ \quad \modeltwo  \\
        $z^{sw,dc}_{\xi re,t-1} - z^{sw,dc}_{\xi re,t} \leq s^{dc}_{t} $&$ \forall \xi re \in \dcswitchtopology \cup \dcZILtopology, \forall t \in 2:T $ \quad \modeltwo \\
        $\sum_{t=1}^{T} s^{dc}_{t} \leq S^{dc,max}$ & $ \forall t \in T $ \quad \modeltwo \\

        \hline
	\end{tabular}
    }
\end{table}

\subsubsection{Topology optimization workflow} \label{subsec:workflow}
\begin{figure*}[h!]
    \centering
    \includegraphics[width=0.8\linewidth]{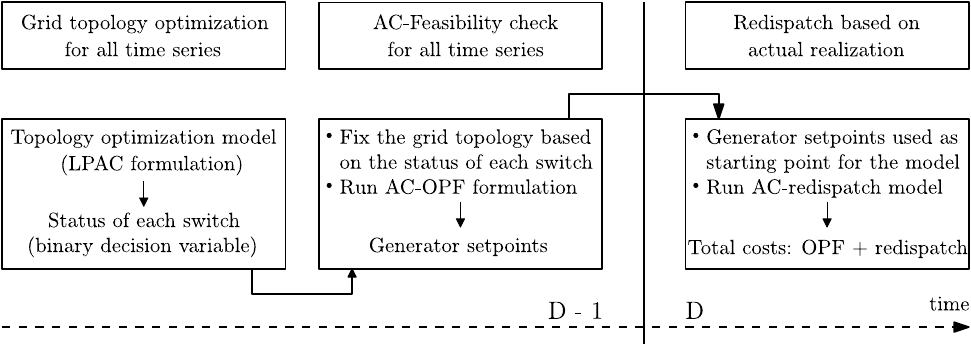}
    \caption{Workflow for the grid topology optimization model presented in the paper}
    \label{fig:workflow}
\end{figure*}

Figure~\ref{fig:workflow} displays the workflow of the grid topology optimization model in the paper. After running Model 2 (and Model 3) using the D-1 RES forecast, the optimized topology is tested for AC-OPF feasibility. At the time of realization D, the RES capacity factor is updated, and a redispatch model is run to compensate for eventual differences in the generator setpoints compared to the D-1 AC-OPF. For example, if the RES generation is lower than expected, a different generator has to increase its generation to ensure that the total generation matches the demand. The redispatch costs for each generator are assumed to be the marginal generation costs.


\subsection{Renewables Energy Sources' uncertainty quantification} \label{subsec:RES}
As introduced in Fig.~\ref{fig:Elia_2024_OFW}, plotting the quarterly difference between the D-1, 11 am wind capacity factor forecast value and the actual measured value from the Belgian Transmission System Operator Elia~\cite{Elia_2024_OFW} for the year 2024, it is still a challenge for TSOs to provide a reliable D-1 forecast for the RES capacity factors in the power grid. This paper starts from this challenge to provide a scenario-based representation of the uncertainty based on the forecast error between the D-1, 11 am wind capacity factor forecast value and the actual measured value from~\cite{Elia_2024_OFW}. While the uncertainty quantification is not the main goal of this paper, it is acknowledged that a considerable amount of work has been dedicated to selecting the best (number of) scenarios to quantify RES uncertainty for short-term grid operations~\cite{Heitsch_2003,Morales_2009,Birge_2011,Papavasiliou_2011,Papavasiliou_2013}. While generally having a high number of scenarios helps to represent the probability distribution of RES uncertainty, the curse of dimensionality leads us to select a limited number of scenarios $W$ to be used in this paper. Results are presented in Section~\ref{sec:results} for 4, 6, and 8 scenarios $w$ per timestep. To generate such scenarios, we follow several steps:
\begin{itemize}
    \item \textit{Analysis of the forecast error between the D-1, 11 am wind capacity factor forecast value and the actual measured value from~\cite{Elia_2024_OFW}}: from Fig.~\ref{fig:Elia_2024_OFW} emerges that:
    \begin{itemize}
        \item The biggest absolute differences take place when the measured value is low and the forecasted value is high.
        \item With high measured capacity factors (above 0.9), the measured value is never severely underestimated. As the difference is contained within 0.0 and 0.6, the forecasted value is rarely higher than the measured one.
        \item With measured capacity factors higher than 0.75, the difference is contained between -0.2 (forecasted value $\geq$ measured value) and 0.6 (measured value $\geq$ forecasted value)
    \end{itemize}
    \item \textit{Computation of the wind uncertainty's probability density function (pdf)}: the pdf of the difference between the measured and the D-1, 11 am forecasted value over the year can be computed with the Kernel Density Estimation method~\cite{Devroye2001} \footnotemark[3]. The resulting pdf is shown in Fig.~\ref{fig:pdf_2024}, its key features are:
    \begin{itemize}
        \item The sharp peak at zero suggests a high frequency of small errors.
        \item The heavy tails suggest that large errors are rare but more frequent than what a Gaussian distribution would suggest
        \item The asymmetry is minimal: the pdf is fairly symmetric around zero, but the length of the tails is different.
    \end{itemize}
    \begin{figure}[h!]
    \centering
    \includegraphics[width=0.65\linewidth]{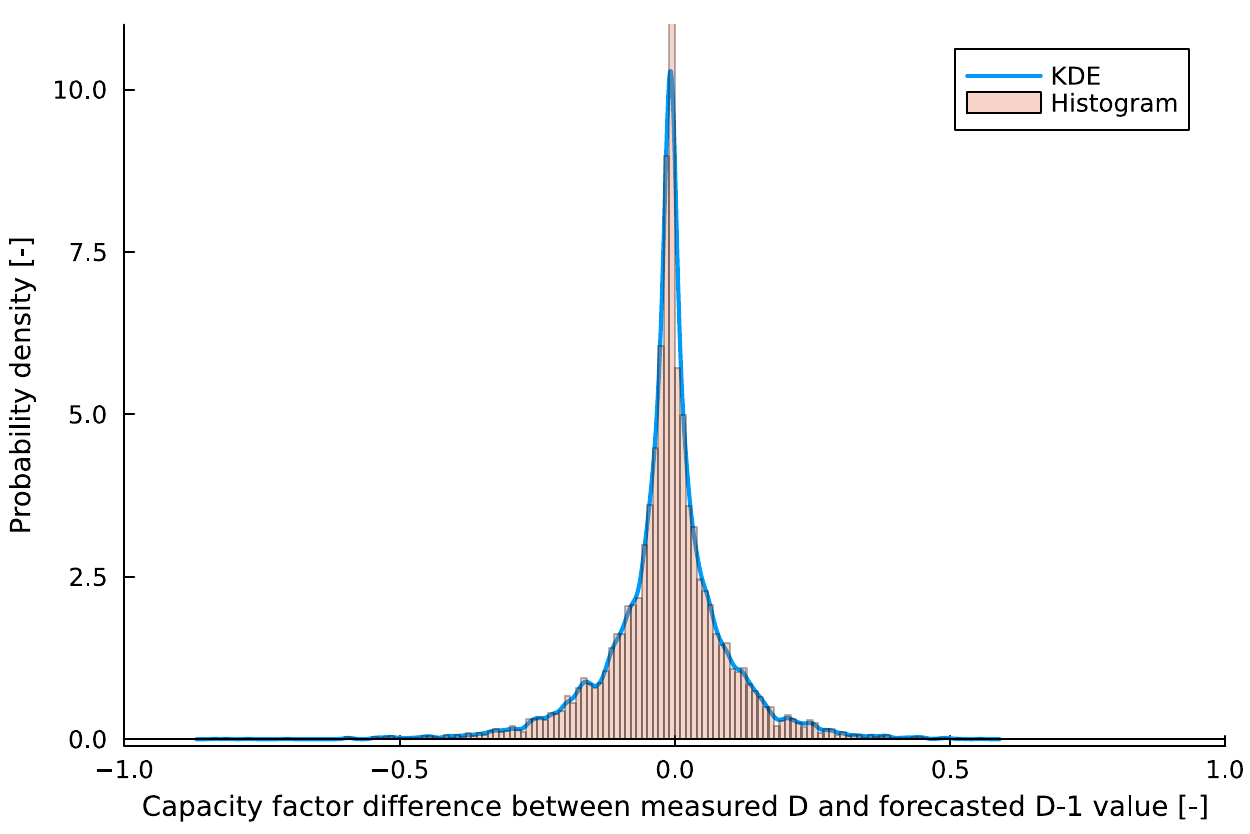}
        \caption{Pdf and related histogram of the difference between the measured and the D-1, 11 am forecasted value over the year 2024 from the EliaOpenData platform~\cite{Elia_2024_OFW}.}
        \label{fig:pdf_2024}
    \end{figure}
    \item \textit{Generating samples and k-means clustering for creating the scenarios}: based on the pdf key features, we fit a Laplace (Double Exponential) distribution to represent the data in Fig.~\ref{fig:pdf_2024} and randomly generate 10$^{5}$ for each timestep. Then, a k-means clustering algorithm~\cite{Jin2010} is used to select $k$ centroids of the randomly selected data, where $k$ equals the number of scenarios $w$ per timestep. The left-hand side of Figure~\ref{fig:time_series} shows the capacity factors for two 24-hours long time series of a forecasted and measured value, and related scenarios. Note that while the scenarios cover a large uncertainty set around the forecasted value, the relative probability of each scenario $w$ is computed based on the pdf shown in previous Fig.~\ref{fig:pdf_2024}.    
\end{itemize}

\footnotetext[3]{The probability density function is computed using the two Julia packages KernelDensity.jl (https://github.com/JuliaStats/KernelDensity.jl) and Distributions.jl (https://github.com/JuliaStats/Distributions.jl).}

\subsection{Input data: test cases \& time series} 
\label{sec:input_data}
\subsubsection{Test cases} \label{subsec:test_cases}

The models formulated in Section~\ref{sec:methodology} are tested on modified versions of the IEEE 30-bus test case~\cite{pglib} and a 50-bus multi-terminal hybrid AC/DC grid, referred to as ``case24\_3zones\_acdc.m'' in~\cite{ergun_optimal_2019}. The number of grid elements for both test cases is summarised in Table~\ref{table:test_cases}. Additional slack generators with higher costs compared to the existing generators are added to each bus in the test cases to model possible load curtailment and for numerical stability. After each grid topology optimization with the LPAC-BS model and AC-OPF feasibility check, the results are compared to see whether load curtailment through these slack generators takes place.
\begin{table}[h!]
\caption{Test cases used to test the grid topology optimization models formulated in this paper.}
\fontsize{9pt}{12pt}\selectfont
\centering
\begin{tabular}{l|cccccc}
\hline
Test case             & \# Generators & \# AC buses & \# DC buses & \#AC/DC conv & \# AC branches & \# DC branches \\
\hline                 
30-bus~\cite{pglib}  & 36 & 30          & -          & -            & 41             & -                   \\
50-bus~\cite{ergun_optimal_2019} & 115 & 50          & 7           & 7             & 77            & 7                   \\
\hline
\end{tabular}
\label{table:test_cases}
\end{table}
Since this paper aims to show the value of grid topology optimization in a $\timesteps$-hours long time period without necessarily finding the most optimal configurations in the test cases, one busbar to be split is selected per test case, namely 6 for the 30-bus test case and 221 for the 50-bus one. Finding metrics to identify the best (combination of) busbars to be split is left for future work. In the 30-bus test case, generation 1 (the largest and cheapest) is assumed to be a wind generator, while in the 50-bus all the cheapest generators (12, with cost lower than 0.14) are selected. At the same time, in the 50-bus case, only the linear term for the objective function M1.1 is kept for each generator and increased by 30\%, while the load of the biggest loads ($>$ 2.0) is increased by 20\%. These actions are performed to obtained a more congested test case.

\subsubsection{Time series} \label{subsec:time_series}

\begin{figure*}[h!]
    \centering
    \begin{subfigure}[b]{0.45\textwidth}
        \centering
        \includegraphics[width=\textwidth]{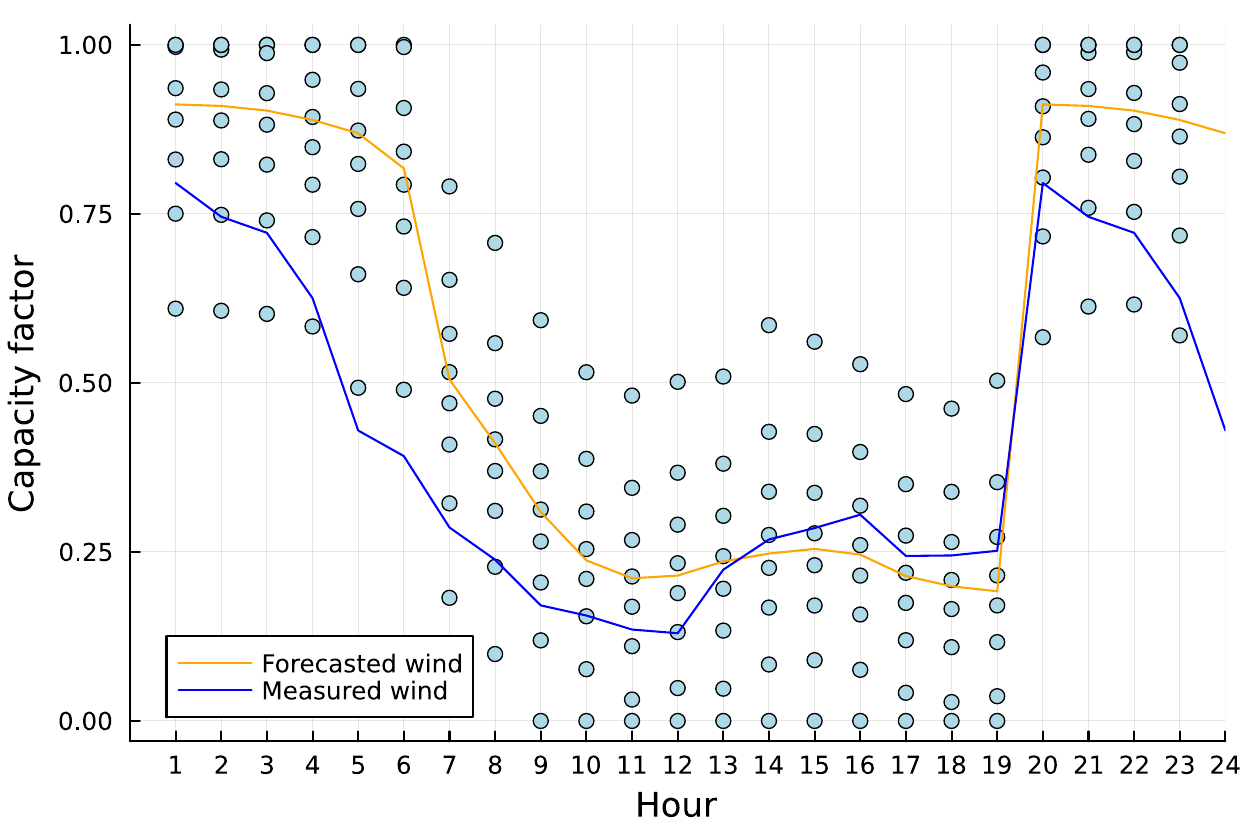}
    \end{subfigure}
    \hfill
    \begin{subfigure}[b]{0.45\textwidth}
        \centering
        \includegraphics[width=\textwidth]{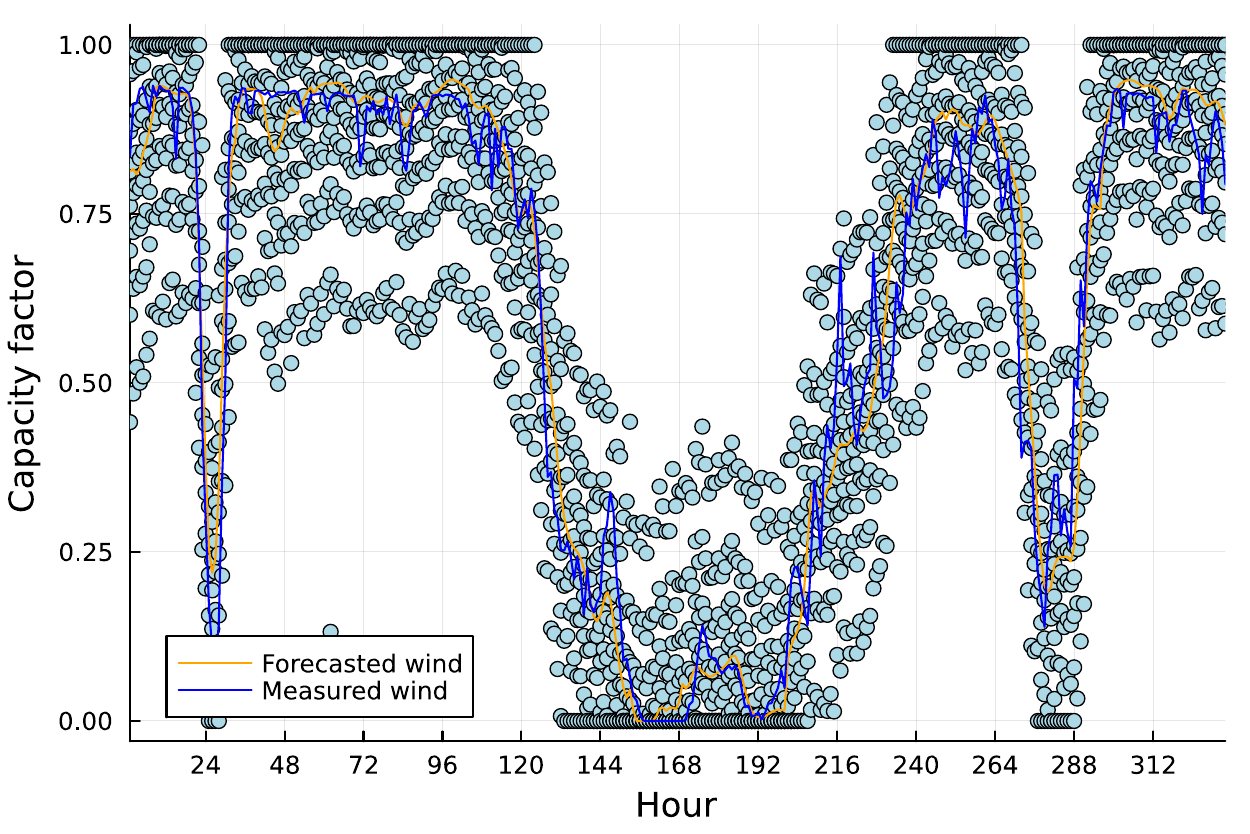}
    \end{subfigure}
    \caption{Time series used for the models developed in the paper. The measured (blue) and forecasted D-1, 11 am (yellow) capacity factors are shown for a day (left) and for two weeks (right). 8 scenarios $w$ (light blue) are used to represent the uncertainty related to the forecasted value.}
    \label{fig:time_series}
\end{figure*}

Two time series are selected to test the grid topology optimization model proposed in the paper. On the left-hand side of Fig.~\ref{fig:time_series}, a 24-hour-long offshore wind time series with two peaks is used to investigate the value of grid topology optimization with high RES variability. Compared to the actual realization, the capacity factor is considerably overestimated at both the beginning and end of the day in D-1, 11 am forecasted value, while it is slightly underestimated from hour 12 to hour 19. The scenarios $w$ computed through the methodology described in Section~\ref{subsec:RES} cover a large spectrum of possible forecast errors. Note that each scenario has its own probability $\omega$, and the sum of all the probabilities $w$ for each timestep $t$ equals 1. Moreover, the right-hand side of Fig.~\ref{fig:time_series} shows 14 days of high wind with a 3-day dip in generation at the end of the first week. Even in this case, there is a considerable forecast error between the forecasted and measured values for several hours in the time series. The two time series will be referred to as \textit{24-hours long} and \textit{14-days long} in the next Section.

The results presented in the next Section are computed using a MacBook Pro with chip M1 Max and 32 GB of memory, using Gurobi~\cite{Gurobi} with MIP gap 0.1\% as optimizer for the MIQCP model and Ipopt~\cite{ipopt} with the linear solver MA97 for the AC-feasibility checks and redispatch parts.


\section{Results and discussion} 
\label{sec:results}

\subsection{AC 30-bus test case}

\subsubsection{24 hours} 
\begin{figure*}[htbp]
  \centering
  \begin{subfigure}[b]{0.32\textwidth}
    \includegraphics[width=\textwidth]{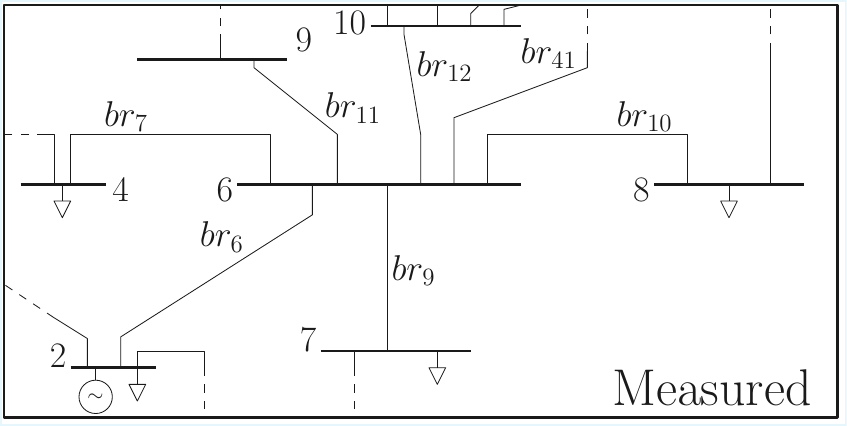}
    \label{fig:sub1}
  \end{subfigure}
  \hfill
  \begin{subfigure}[b]{0.32\textwidth}
    \includegraphics[width=\textwidth]{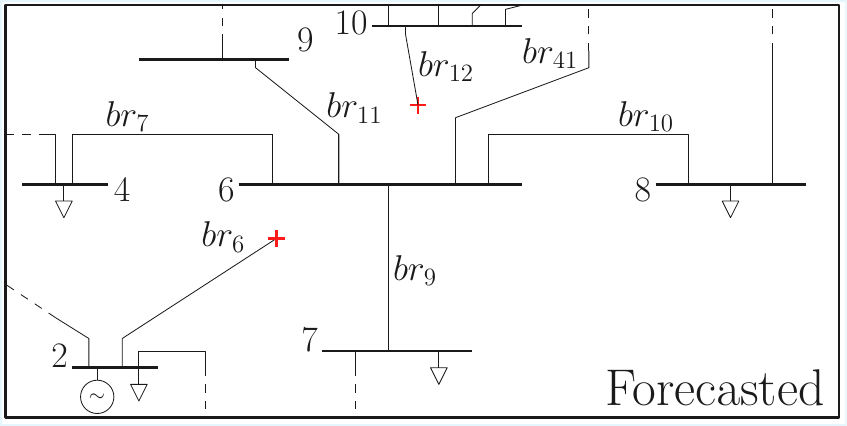}
    \label{fig:sub2}
  \end{subfigure}
  \hfill
  \begin{subfigure}[b]{0.32\textwidth}
    \includegraphics[width=\textwidth]{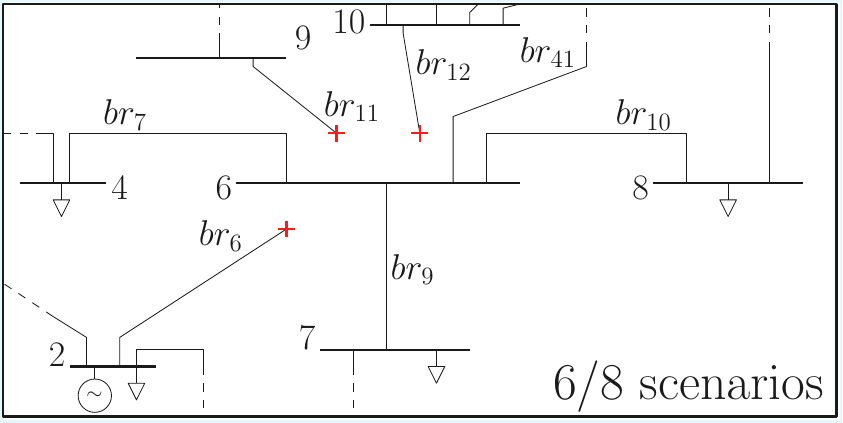}
    \label{fig:sub3}
  \end{subfigure}
  \caption{Optimized topology of busbar 6 for the \textit{Measured} (left), \textit{Forecasted} (center), and \textit{6/8 scenarios $\scenarios$} (right) when only one topology is allowed in the 24-hour simulation. }
  \label{fig:topologies}
\end{figure*}

Based on the 24-hour time series from Fig.~\ref{fig:time_series}, we analyze how an overestimation of the D capacity factors influences the optimized grid topology. For this test case, it is observed that topological actions are performed only when the wind capacity factors exceed $0.8$, as grid congestion and wind curtailment occur at this threshold. Therefore, optimizing the topology based on an overestimation of the wind resource leads to an optimal topology for the D-1 forecast, but sub-optimal for the D measured values. Consequently, as shown in Table~\ref{table:results_30}, the total generation costs (sum of the AC-OPF feasibility check and redispatch models) for the \textit{Forecasted} time series are considerably higher than the topology based only on the D \textit{Measured} values, which has perfect foresight on the wind realization and is therefore the one with the lowest costs. In this particular case, no topological actions need to be performed, as indicated by the \textit{Measured} column in Table~\ref{table:results_30}, and shown on the left-hand side of Figure~\ref{fig:topologies}.  

When uncertainty on the RES forecast is considered using the methodology described in Section~\ref{subsec:RES}, the total costs of the simulations with 6 (\textit{6 $\scenarios$}) and 8 scenarios (\textit{8 $\scenarios$}) are lower compared to the \textit{Forecasted} results based only on the D-1 forecast. Differently from the \textit{Forecasted} case, the scenarios-based topology optimization models do lead to lower costs than their respective OPF simulations too. For the \textit{One topology} simulations, the optimal topologies in the center and right-hand side of Figure~\ref{fig:topologies} are also different, with OTS performed in branches $6$ and $12$ for the \textit{Forecasted} case, and on branches $6$, $11$, and $12$ for $6$/$8$ scenarios $\scenarios$. 
Note that in the simulation with 4 scenarios (\textit{4 $\scenarios$}), on the other hand, the total costs are higher than the \textit{Forecasted} ones, suggesting that 4 scenarios are not enough to guarantee an acceptable uncertainty representation. 

\begin{table}[h!]
\caption{Increase in generation costs [\%] for a 24-hour simulation compared to the results of the \textit{Measured} OPF value. Results are shown for the measured capacity factors at time D (\textit{Measured}), the capacity factors of the D-1, 11 am forecast (\textit{Forecasted}), and simulations with 4, 6 and 8 scenarios $\scenarios$. The grid topology is optimized hourly (Hourly BS), once in 24 hours (\textit{One topology}), or allows up to one (\textit{Up to one switching action}) or two switching actions (\textit{Up to two switching actions}) for each time series. The generation costs are the sum of the D-1 AC-OPF feasibility check and redispatch models based on the measured D capacity factor. If the wind capacity factor is above a certain threshold, renewable curtailment occurs. Since the wind capacity factor does not go over the threshold for the measured D time series, but it does for the forecasted D-1 one, there is no value of topological actions for the \textit{Measured} case.}
\centering
{\fontsize{9.0pt}{11pt}\selectfont
\begin{tabular}{c|ccccc}
\hline
Simulation & Measured & Forecasted  & 4 $\scenarios$ & 6 $\scenarios$ & 8 $\scenarios$ \\
\hline
OPF         & 0 & 10.2 &	9.2	& 13.3	& 11.8 \\
Hourly BS	 & 0 & 14    & 15.9	& 4.5 & 3.8 \\
One topology & 0 & 14.2	 & 17.4	& 5.0 & 4.5 \\
Up to one switching action    & 0 & 14.2 & 16.1	& 4.3 & 1.8 \\
Up to two switching actions & 0 & 13.9	 & 15.2	& 4.3 & 1.8 \\
\hline
\end{tabular}
}
\label{table:results_30}
\end{table}

Regarding the other simulations, as expected, leaving the freedom to operate one/two switching actions leads to lower total generation costs compared to fixing the topology once (\textit{One topology}) in the 24 hours. While the \textit{Hourly-BS} is the cheapest for the LPAC-BS model, the models allowing switching actions score lower total generation costs compared to the \textit{Hourly-BS}. This fact can be explained by the fact that i) LPAC-BS model is an approximation of the ``AC'' model, thus the benefit brought by LPAC-BS topologies is not necessarily directly reflected in the AC-feasibility checks, and ii) the benefit brought by optimizing the grid topology is not considerable in this case, thus there might be simmetry/low differences brought by selecting one or the other topology. Nevertheless, these results for this specific test case and time series show how including the forecast uncertainty via 6 and 8 scenarios $\scenarios$ leads to lower total costs compared to basing only on the forecasted D-1 capacity factors. The results demonstrate that the quality of the RES forecast significantly influences the economic benefits brought by the optimized topology. Therefore, grid operators should evaluate the efficiency in performing topological actions depending on the quality of the forecast and on the characteristics of the power grid.

\subsubsection{14 days}
In this section, we demonstrate the economic benefits of performing grid topology optimization over a longer period (14 days) compared to the previous subsection. Besides normal grid operations, the proposed models are also relevant for maintenance planning, when parts of the grid or multiple network elements are taken out of service, requiring adjustments to the usual grid topology.

The wind time series shown on the right-hand side of Fig.~\ref{fig:time_series} is used as input for the wind generator in the test case, previously introduced in Section~\ref{subsec:test_cases}. Similar to Table~\ref{table:results_30}, Table~\ref{table:results_30_14_days} shows the cost increase compared to the ~\textit{Measured} OPF for the different simulations and time series, but also adds a column with the cost increase (or decrease) of each simulation compared to the OPF costs of the same time series. For example, in the perfect foresight \textit{Measured} case, the reduction in total generation costs ranges from 1.67\% (\textit{One topology}) to 1.89\% (\textit{Hourly BS}). In the \textit{Forecasted} time series, the simulations see a cost increase compared to the~\textit{Measured} OPF on the left column, and a cost decrease ranging between 3.68\% and 3.76\% compared to the \textit{Forecasted} OPF on the right column. 

\begin{figure*}[h!]
    \centering
    \begin{subfigure}[b]{0.48\textwidth}
        \centering
        \includegraphics[width=\textwidth]{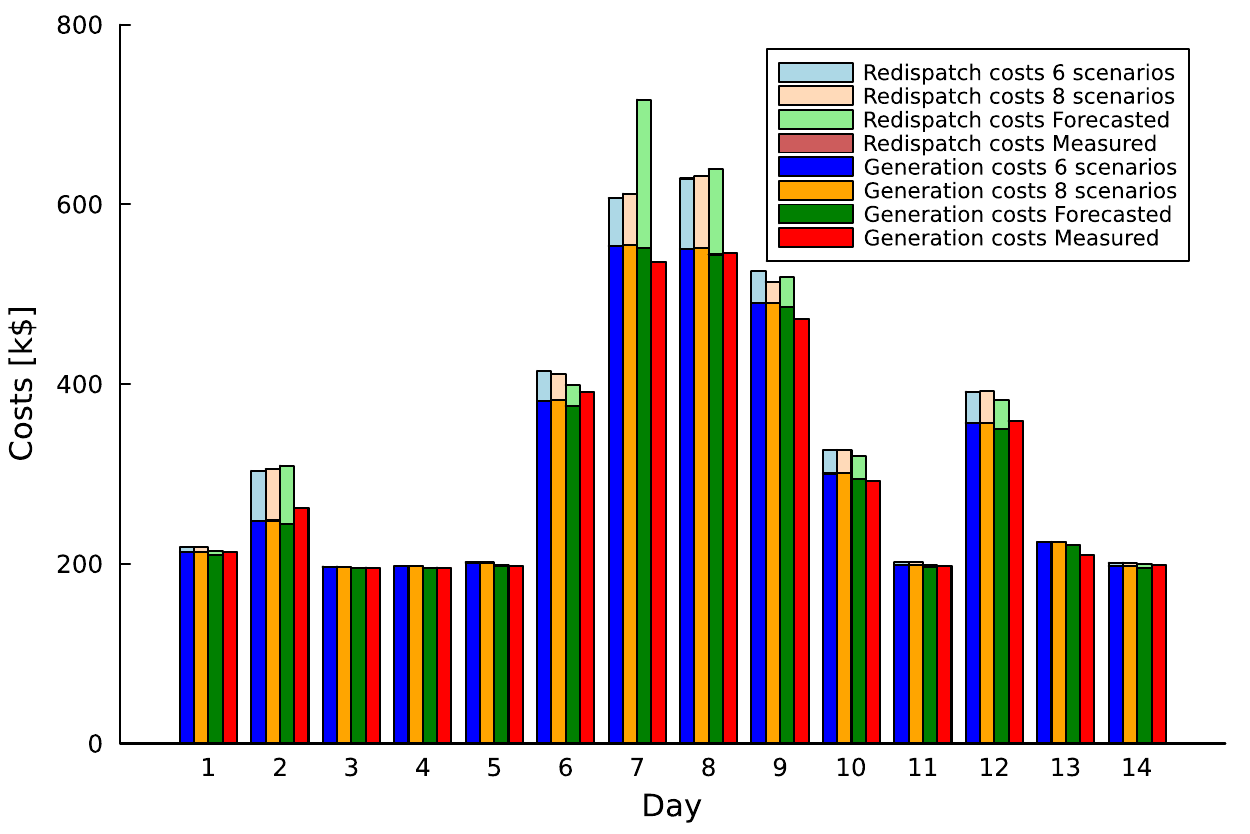}
    \end{subfigure}
    \hfill
    \begin{subfigure}[b]{0.48\textwidth}
        \centering
        \includegraphics[width=\textwidth]{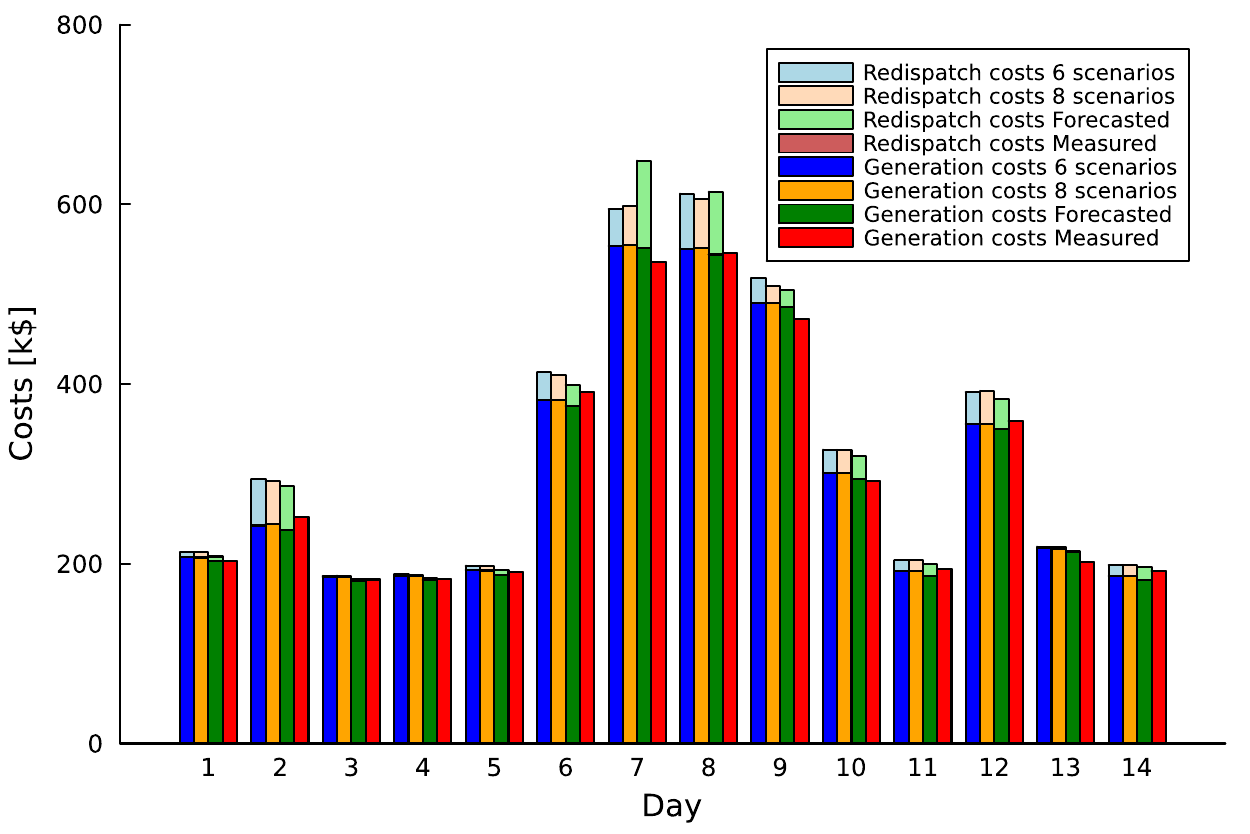}
    \end{subfigure}
    \caption{Daily \textit{Optimal power flow} (left) and \textit{One topology} (right) generation and redispatch costs for four time series: \textit{$6 \scenarios$}, \textit{$8 \scenarios$}, \textit{Forecasted} and \textit{Measured}.}
    \label{fig:Gen_and_redispatch_cost_comparison}
\end{figure*}

We note a similar trend and comparable benefits as the \textit{Measured} and \textit{Forecasted} cases also for the \textit{6 $\scenarios$} and \textit{8 $\scenarios$} time series in Table~\ref{table:results_30_14_days}, with all simulations being cheaper than the total costs of the OPF of their respective time series. Especially for the \textit{Forecasted} time series, the benefits brought by optimizing the grid topology are more than 3.6\% for all the simulations, but the total OPF costs are high due to the forecast error in the D-1 capacity factors. This fact is confirmed by the left-hand side of Fig.~\ref{fig:Gen_and_redispatch_cost_comparison}, with the \textit{Forecasted} time series having considerably higher average and peak redispatch costs compared to the scenario-based time series. Even when the \textit{Forecasted} generation cost is smaller, it incurs a larger redispatch cost. As a result, quantifying the uncertainty through $6$-$8$ scenarios $\scenarios$ leads to a more reliable D-1 forecast for the OPF simulations in this test case. For the other simulations, on the other hand, the results of the \textit{Forecasted}, \textit{6 $\scenarios$} and \textit{8 $\scenarios$} are comparable, as displayed by the right-hand side of Fig.~\ref{fig:Gen_and_redispatch_cost_comparison}, where the redispatch costs are lower than the OPF counterpart on the right-hand side of Fig.~\ref{fig:Gen_and_redispatch_cost_comparison}. In addition, allowing a limited number of switching actions per day reduces the total costs in the \textit{Measured} and \textit{Forecasted} time series compared to the \textit{One topology} simulations, while they do not lead to cost reductions for the scenario-based simulations.

\begin{table*}[h!]
\caption{i) Increase in total costs (generation and redispatch) [\%] compared to the \textit{Measured} OPF value computed using the D measured offshore wind capacity, and ii) increase in total costs (generation and redispatch) [\%] compared to the respective time series' OPF value for a 14-day simulation. Results are shown for the measured capacity factors at time D (\textit{Measured}), the capacity factors of the D-1, 11 am forecast (\textit{Forecasted}), and simulations with 6 and 8 scenarios $\scenarios$. The grid topology is optimized hourly (\textit{Hourly BS}), once every 24 hours (\textit{One topology}), or allowing up to one (\textit{Up to one switching action}) or two switching actions (\textit{Up to two switching actions}) for each time series. The total costs are the sum of the D-1 AC-OPF feasibility check and the redispatch models based on the measured D capacity factor.}
\centering
{\fontsize{7pt}{10pt}\selectfont
\begin{tabular}{c|c|cc|cc|cc}
\cline{1-8}
 Time series $\rightarrow$& Measured & \multicolumn{2}{c|}{Forecasted}  & \multicolumn{2}{c|}{6 $\scenarios$} & \multicolumn{2}{c}{8 $\scenarios$} \\
\hline
  & \% increase    &  \% increase    & \% increase  & \% increase    & \% increase & \% increase    & \% increase \\
 Simulation $\downarrow$ & wrt Measured  & wrt Measured  & wrt Forecasted  & wrt Measured  & wrt 6 $\scenarios$  & wrt Measured  & wrt 8 $\scenarios$ \\
 & OPF  & OPF &  OPF & OPF & OPF & OPF & OPF \\
\hline
OPF          & 0       &    10.38   &	0	    & 8.73 &	0	& 8.59 & 0 \\
Hourly BS	 &	-1.89  &	6.18 &	- 3.81	& 6.63 &	- 1.94	& 6.44 & - 1.97 \\
One topology & -1.67   &	6.33 &	- 3.68	& 6.80 &	- 1.78	& 6.46 & - 1.96 \\
Up to one    & \multirow{2}{*}{-1.80} & \multirow{2}{*}{6.23} & \multirow{2}{*}{-3.76} 	& \multirow{2}{*}{6.81} & \multirow{2}{*}{-1.78} & \multirow{2}{*}{6.47} & \multirow{2}{*}{-1.96} \\
switching action & 	 &  & &  &	  	&  &  \\
Up to two 	& \multirow{2}{*}{-1.80}  & \multirow{2}{*}{6.50} & \multirow{2}{*}{-3.76}	& \multirow{2}{*}{6.81} & \multirow{2}{*}{-1.78} & \multirow{2}{*}{6.47} & \multirow{2}{*}{-1.96} \\
switching actions &	  & &  &  	&  & &  \\
\hline
\end{tabular}
}
\label{table:results_30_14_days}
\end{table*}

\begin{figure}[h!]
    \centering
    \includegraphics[width=0.65\textwidth]{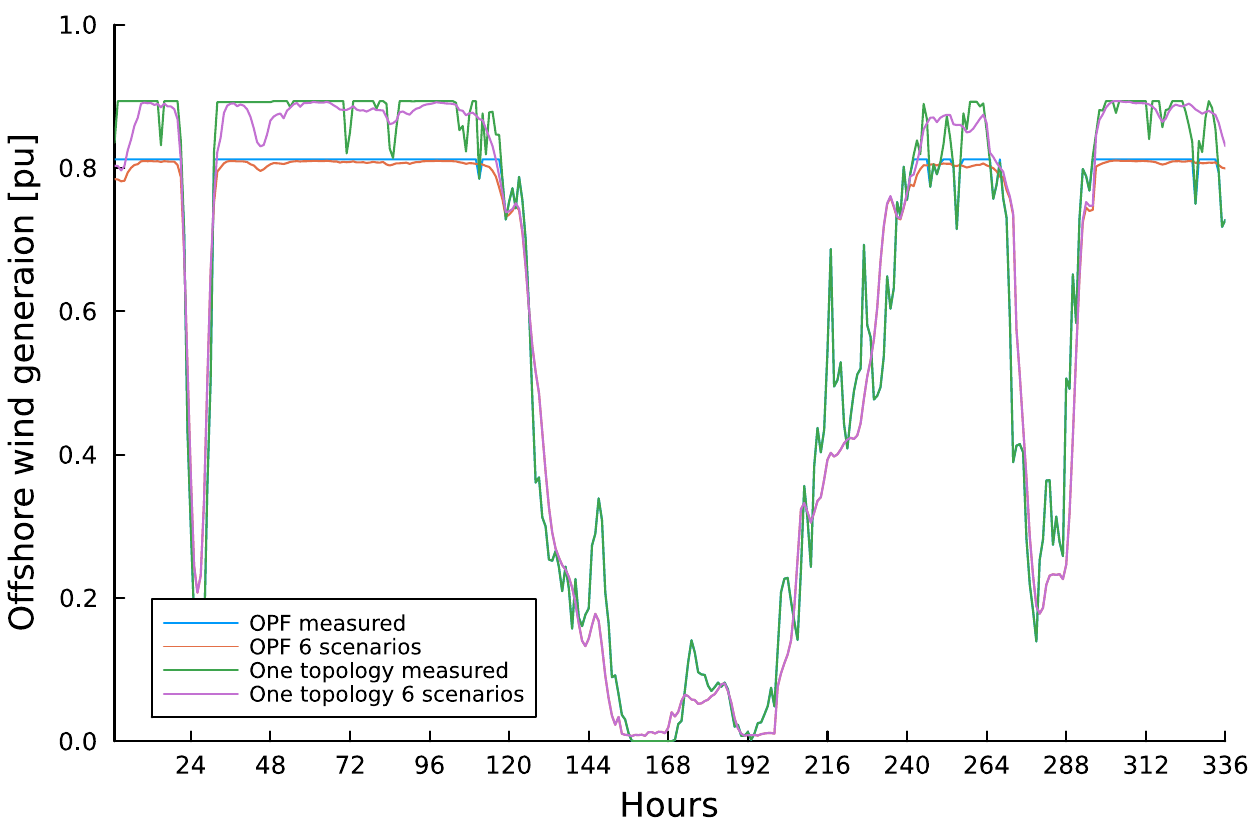}
    \caption{Offshore wind generation for the optimal power flow simulations and while daily optimizing the topology (One topology) in the selected 14 days. The actual measured capacity factors (measured) and the 6-scenario-based renewable uncertainty quantification based on the day-ahead forecast (6 scenarios $\scenarios$) are used as input for the wind generators in the test case.}
    \label{fig:wind_gen_14_days}
\end{figure}

Moreover, Fig.~\ref{fig:wind_gen_14_days} shows how optimizing the topology in \textit{One topology} leads to less RES curtailment compared to the OPF simulations for the \textit{Measured} and \textit{6 $\scenarios$} time series, and therefore to lower total costs. Nevertheless, while the wind generation for the \textit{Measured} and \textit{$6  \scenarios$} in Fig.~\ref{fig:wind_gen_14_days} are comparable for the OPF simulations, we still notice a gap between the wind generation in the \textit{One topology} cases. Specifically, the peaks in generation between days $5$ and $10$ are not well-captured by the generated scenarios. Therefore, more precise forecast models can result in a better representation of the wind generation, and therefore less redispatch costs. While providing a better RES forecast than the \textit{Forecasted} time series was not the goal of this work, the economic value of having accurate forecasts is clear, and a more accurate scenario selection methodology will be investigated in future work.

%

Regarding the simulation time for both the 24 hours and 14 days time series, they range from 0.34~s per hour (8.16~s per day) for the \textit{Hourly BS} to 6550.26~s for the \textit{Up to one switching action} with 8 scenarios $\scenarios$. The AC-feasibility checks and redispatch simulations have negligible computational times compared to the LPAC-BS model. A complete overview of the computational times for each simulation and time series is added in Appendix A at the end of the paper.

\subsection{Hybrid AC/DC 50-bus test case, 14 days}

We extend the analysis on the 30-bus test case from the previous Section to the 50-bus hybrid AC/DC grids introduced earlier in Table~\ref{table:test_cases}. To keep the text concise, we focus mainly on the economic benefit brought by the presented models to the total grid operation costs. Therefore, only the results for the 14 days time series are presented. The main trends seen with the 30-bus test case are confirmed for this 50-bus test case, as shown in Table~\ref{table:results_50_14_days}. First, in the \textit{Measured} time series, the grid topology optimization models lead to a reduction in generation costs of 0.02-0.04\%, which is smaller than for the 30-bus test case, and in the vicinity of the 0.1\% MIP gap. For the \textit{Forecasted}, \textit{6~$\scenarios$} and \textit{$8~\scenarios$} time series, the reduction in total costs compared to their respective OPF is considerable, ranging from a minimum of 0.31~\% in \textit{6~$\scenarios$} to a maximum of 5.88\% for the \textit{Hourly BS} of \textit{Forecasted}. As in previous Table~\ref{table:results_30_14_days}, the possibility of switching actions does not lead to additional cost reductions for the \textit{6~$\scenarios$} and \textit{$8~\scenarios$} time series, while it further reduces the costs for the \textit{Forecasted} time series by 0.08\%. As the generated scenarios~$\scenarios$ already cover an extended range of forecast errors, the effect of additional switching actions does not bring additional economic benefits with this input data. Despite this fact, in this test case, using a scenario-based representation of the RES forecast uncertainty results in lower total costs compared to the~\textit{Forecasted} time series for all the simulations, confirming the effectiveness of the proposed methodology. These results are comparable to what was observed for case 30-bus in the grid topology optimization models of the 24 hours simulation, and the OPF of the 14 days simulation. In this latter simulation, the grid topology optimization models of the \textit{6$\scenarios$~/~8$\scenarios$} time series had sensibly higher costs than the \textit{Forecasted} one. Therefore, as introduced earlier in the text, further refining the RES uncertainty quantification possibly guarantees lower costs than the~\textit{Forecasted} time series for a larger set of time instances.

\begin{table*}[h!]
\caption{i) Increase in total costs (generation and redispatch) [\%] compared to the \textit{Measured} OPF value computed using the D measured offshore wind capacity, and ii) increase in total costs (generation and redispatch) [\%] compared to the respective time series' OPF value for a 14-day simulation. Results are shown for the measured capacity factors at time D (\textit{Measured}), the capacity factors of the D-1, 11 am forecast (\textit{Forecasted}), and simulations with 6 and 8 scenarios $\scenarios$. The grid topology is optimized hourly (\textit{Hourly BS}), once every 24 hours (\textit{One topology}), or allowing up to one (\textit{Up to one switching action}) or two switching actions (\textit{Up to two switching actions}) for each time series. The total costs are the sum of the D-1 AC-OPF feasibility check and the redispatch models based on the measured D capacity factor.}

\centering
{\fontsize{7pt}{10pt}\selectfont
\begin{tabular}{c|c|cc|cc|cc}
\cline{1-8}
 Time series $\rightarrow$& Measured & \multicolumn{2}{c|}{Forecasted}  & \multicolumn{2}{c|}{6 $\scenarios$} & \multicolumn{2}{c}{8 $\scenarios$} \\
\hline
  & \% increase    &  \% increase    & \% increase  & \% increase    & \% increase & \% increase    & \% increase \\
 Simulation $\downarrow$ & wrt Measured  & wrt Measured  & wrt Forecasted  & wrt Measured  & wrt 6 $\scenarios$  & wrt Measured  & wrt 8 $\scenarios$ \\
 & OPF  & OPF &  OPF & OPF & OPF & OPF & OPF \\
\hline
OPF          & 0         & 10.71   &	0	    & 5.06 &	0	& 8.59 & 0 \\
Hourly BS	 & -0.04     &	4.20 & -5.88	& 3.68 &	-1.32	& 2.65 & -5.47 \\
One topology & -0.02     &	5.10 & -5.07	& 4.74 &	-0.31	& 4.88 & -3.41 \\
Up to one    & \multirow{2}{*}{-0.02} & \multirow{2}{*}{5.03} & \multirow{2}{*}{-5.13} 	& \multirow{2}{*}{4.74} & \multirow{2}{*}{-0.31} & \multirow{2}{*}{4.88} & \multirow{2}{*}{-3.41} \\
switching action &  &  & &  &	  	&  &  \\
Up to two 	& \multirow{2}{*}{-0.02} & \multirow{2}{*}{5.03} & \multirow{2}{*}{-5.13}	& \multirow{2}{*}{4.74} & \multirow{2}{*}{-0.31} & \multirow{2}{*}{4.88} & \multirow{2}{*}{-3.41} \\
switching actions &  & &  &  	&  & &  \\
\hline
\end{tabular}
}
\label{table:results_50_14_days}
\end{table*}

Regarding the average simulation time for these 14 days simulations, they range from 0.33~s per hour (7.92~s per day) for the \textit{Hourly BS} to 10397.70~s for the \textit{Up to two switching action}, \textit{Forecasted}. The AC-feasibility checks and redispatch simulations have negligible computational times compared to the LPAC-BS model. A complete overview of the computational times for each simulation and time series is added in Appendix A at the end of the paper.

\subsubsection{Comment on DC busbar splitting}
Note that throughout the paper, we focused only on busbar splitting of AC busbars, while in Model 1 DC busbar splitting is also added. While DC busbar splitting does possibly lead to an economic benefit in grid topology optimization models~\cite{Bastianel}, it is still an emerging concept that requires, among others, the coordination between AC/DC converter controls and DC protections. In addition, the currently available test cases still do not have extended mult-terminal (and meshed) DC grids, which would be the most logical application of DC busbar splitting. For these reasons, DC busbar splitting is still presented in Model 1 as it can be combined with AC busbar splitting, and its application to extended DC grids is left for future work.


\section{Conclusion and future work} \label{sec:conclusion}
This paper introduces for the first time a multistep and stochastic grid topology optimization model that combines busbar splitting and optimal transmission switching for both AC and hybrid AC/DC grids. The renewable energy sources' uncertainty is represented with a scenario-based approach. The scenarios are based on the forecast error distribution between the day-ahead and measured values for offshore wind in Belgium in 2024. 

For the 24 hours simulation, we show the need for accurate renewable forecasts, or more advanced methods to quantify renewable uncertainty, when using steady-state grid topology optimization models. The results for the 14 days prove how the proposed grid topology optimization models do lead to lower total (AC-feasibility check and redispatch) costs compared to an optimal power flow simulation with the original topology. While optimizing the topology hourly does lead to the lowest costs for most of the simulations, we show how selecting one topology per day, or allowing up to one/two switching actions per day, have comparable results to the hourly optimization model. Therefore, considerable reductions in total costs can still be achieved without involving topological changes every hour, reducing the complexity of such topological actions on the power grid. Moreover, quantifying renewable uncertainty through the scenario-based methodology we propose helps to mitigate the redispatch costs, especially in the optimal power flow simulation, leading to lower total costs compared to the results obtained by basing only on the day-ahead forecast.
For the grid topology optimization models, the reduction in costs for the day-ahead and scenario-based time series with respect to the measured values leads to better or at least comparable (14 days case) results than relying on a single forecasted value.

Future work will include the addition of security constraints to the model in order to test the optimized topologies in case of N-1 contingencies happening in the grid. Due to the size of the problem, mathematical decompositions will be explored. In addition, methods to select (combinations) of the most relevant busbars to be split in the grid will be developed for increasing the computational efficiency of the topology optimization problem in general. Lastly, the proposed methodologies will be tested on real-life inspired test cases for hybrid AC/DC grids to quantify the value of grid topology optimization based on real data, and propose optimal topologies for new multi-terminal DC grids.


\section*{Acknowledgement}
This paper has received support from the Belgian Energy Transition Fund, FOD Economy, project DIRECTIONS, and from a Research Foundation – Flanders (Belgium) (FWO) travel grant for Giacomo Bastianel's research visit at the University of Wisconsin-Madison.

\section*{Declaration of generative AI and AI-assisted technologies in the writing process}

During the preparation of this work, the author(s) used ChatGPT to improve text clarity and assist with coding. All content generated with the help of this tool was reviewed and edited by the author(s), who take full responsibility for the final version of the publication..

\bibliographystyle{elsarticle-num} 
\bibliography{Bibliography}

\begin{thebibliography}{10}
\expandafter\ifx\csname url\endcsname\relax
  \def\url#1{\texttt{#1}}\fi
\expandafter\ifx\csname urlprefix\endcsname\relax\def\urlprefix{URL }\fi
\expandafter\ifx\csname href\endcsname\relax
  \def\href#1#2{#2} \def\path#1{#1}\fi

\bibitem{EU_1}
{European Commission}, Renewable energy directive, [Online] Available: https://energy.ec.europa.eu/topics/renewable-energy/renewable-energy-directive-targets-and-rules\_en. (accessed July 3, 2025) (2023).

\bibitem{EU_2}
{European Commission}, {REPowerEU Affordable, secure and sustainable energy for Europe}, [Online]. Available: https://commission.europa.eu/strategy-and-policy/priorities-2019-2024/european-green-deal/repowereu-affordable-secure-and-sustainable-energy-europe\_en (accessed July 3, 2025) (2022).

\bibitem{IRENA}
IRENA, {World Energy Transitions Outlook 2023: 1.5°C Pathway}, [Online] Available: https://www.irena.org/Publications/2023/Jun/World-Energy-Transitions-Outlook-2023. (accessed July 3, 2025) (2023).

\bibitem{GSC}
{Global Solar Council}, {Global Solar Council announces 2 terawatt milestone achieved for solar}, [Online] Available: https://www.globalsolarcouncil.org/news. (accessed July 3, 2025) (2024).

\bibitem{WWEA}
{World Wind Energy Association}, {WWEA Annual Report 2023}, [Online] Available: https://wwindea.org/ss-uploads/media/2024/3/1711538106-40ab83f2-3e01-4c0a-9d28-e0a21bff72e6.pdf. (accessed July 3, 2025) (2024).

\bibitem{Elia_2024_OFW}
OpenDataElia, {Wind power production estimation and forecast on Belgian grid (Near real-time)}, [Online] Available: https://opendata.elia.be/explore/dataset/ods086/information/. (accessed July 7, 2025) (2025).

\bibitem{IEA_2023}
{International Energy Agency}, Electricity grids and secure energy transitions, [Online] Available: https://www.iea.org/reports/electricity-grids-and-secure-energy-transitions. (accessed July 3, 2025) (2023).

\bibitem{JRC_2024}
{European Commission, Joint Research Centre}, Redispatch and congestion management, [Online] Available: https://data.europa.eu/doi/10.2760/853898, JRC137685. (accessed July 3, 2025) (2024).

\bibitem{ACER_congestion}
{Agency for the Cooperation of Energy Regulators}, {Transmission capacities for cross-zonal trade of electricity and congestion management in the EU}, [Online] Available: https://www.acer.europa.eu/sites/default/files/documents/Publications. (accessed July 6, 2025) (2025).

\bibitem{Congestion_US}
N.~Shreve, J.~Selker, Z.~Zimmerman, R.~Gramlich, 2023 transmission congestion report, [Online] Available: https://gridstrategiesllc.com/wp-content/uploads/Grid-Strategies\_2023-Transmission-Congestion-Report.pdf (accessed July 6, 2025) (2024).

\bibitem{DE_COSTS}
{ENTSO-E transparency platform}, Costs of congestion management, [Online] Available: https://transparency.entsoe.eu/congestion-management. (accessed July 6, 2024).

\bibitem{SMARD}
{SMARD.de}, Congestion management, [Online] Available: https://www.smard.de/page/en/topic-article/5790/214060/congestion-management (accessed July 6, 2025) (2025).

\bibitem{IEA_2025}
{International Energy Agency}, Electricity 2025: Analysis and forecast to 2027, [Online] Available: https://iea.blob.core.windows.net/assets/0f028d5f-26b1-47ca-ad2a-5ca3103d070a/Electricity2025.pdf. (accessed July 3, 2025) (2024).

\bibitem{IEA_2025_constraints}
{International Energy Agency}, Building the future transmission grid - strategies to navigate supply chain challenges, [Online] Available: https://energy.ec.europa.eu/topics/renewable- energy/renewable-energy-directive-targets-and-rules\_en. (accessed July 6, 2025) (2025).

\bibitem{NEUKIRCH_Grid_acceptance}
M.~Neukirch, Grinding the grid: Contextualizing protest networks against energy transmission projects in southern germany, Energy Research \& Social Science 69 (2020) 101585.
\newblock \href {https://doi.org/https://doi.org/10.1016/j.erss.2020.101585} {\path{doi:https://doi.org/10.1016/j.erss.2020.101585}}.

\bibitem{nrao}
{CCR Core TSO Cooperation}, Explanatory document to the \uppercase{C}ore \uppercase{C}apacity \uppercase{C}alculation \uppercase{R}egion methodology for common provisions for regional operational security coordination in accordance with \uppercase{A}rticle 76 of \uppercase{C}ommission \uppercase{R}egulation (\uppercase{E}\uppercase{U}) 2017/1485 (2019).

\bibitem{EPRI}
{EPRI}, Transmission topology optimization, state-of-the-art white paper: A get set white paper, [Online] Available: https://www.epri.com/research/products/000000003002031445 (accessed July 6, 2025) (2025).

\bibitem{ENTSO_E_CCR}
{ENTSO-E}, {Capacity calculation in Core CCR}, [Online] Available: https://www.entsoe.eu/bites/ccr-core/explained/. (accessed July 6, 2025).

\bibitem{Bastianel}
G.~Bastianel, M.~Vanin, D.~{Van Hertem}, H.~Ergun, {Optimal Transmission Switching and Busbar Splitting in Hybrid AC/DC Grids}, arXiv:2412.00270 (2024).
\newblock \href {http://arxiv.org/abs/2412.00270} {\path{arXiv:2412.00270}}.

\bibitem{Fisher2008}
E.~B. Fisher, R.~P. O'Neill, M.~C. Ferris, Optimal transmission switching, IEEE Transactions on Power Systems 23~(3) (2008) 1346--1355.
\newblock \href {https://doi.org/10.1109/TPWRS.2008.922256} {\path{doi:10.1109/TPWRS.2008.922256}}.

\bibitem{Hedman2009}
K.~Hedman, R.~O'Neill, E.~Fisher, S.~Oren, Optimal transmission switching with contingency analysis, in: IEEE PES General Meeting, 2010.
\newblock \href {https://doi.org/10.1109/PES.2010.5589437} {\path{doi:10.1109/PES.2010.5589437}}.

\bibitem{Morsy2022}
B.~Morsy, A.~Hinneck, D.~Pozo, J.~Bialek, Security constrained {OPF} utilizing substation reconfiguration and busbar splitting, EPSR 212 (2022) 108507.
\newblock \href {https://doi.org/10.1016/j.epsr.2022.108507} {\path{doi:10.1016/j.epsr.2022.108507}}.

\bibitem{Hinneck2022}
A.~Hinneck, D.~Pow, Solving security-constrained transmission switching using column-and-constraint substitution, in: 13th Mediterranean Conference on Power Generation, Transmission, Distribution and Energy Conversion (MEDPOWER 2022), Vol. 2022, 2022, pp. 267--272.
\newblock \href {https://doi.org/10.1049/icp.2023.0004} {\path{doi:10.1049/icp.2023.0004}}.

\bibitem{MORSY_2025}
B.~Morsy, M.~Deakin, A.~Anta, J.~Cremer, Corrective soft bus-bar splitting for reliable operation of hybrid ac/dc grids, International Journal of Electrical Power \& Energy Systems 169 (2025) 110792.
\newblock \href {https://doi.org/https://doi.org/10.1016/j.ijepes.2025.110792} {\path{doi:https://doi.org/10.1016/j.ijepes.2025.110792}}.

\bibitem{Heidarifar2016}
M.~Heidarifar, H.~Ghasemi, A network topology optimization model based on substation and node-breaker modeling, IEEE Transactions on Power Systems 31~(1) (2016) 247--255.
\newblock \href {https://doi.org/10.1109/TPWRS.2015.2399473} {\path{doi:10.1109/TPWRS.2015.2399473}}.

\bibitem{Baker2021}
K.~Baker, {Solutions of DC OPF are Never AC Feasible} (2020).
\newblock \href {http://arxiv.org/abs/1912.00319} {\path{arXiv:1912.00319}}.

\bibitem{Heidarifar2021}
M.~{Heidarifar, et al.}, {An optimal transmission line switching and bus splitting heuristic incorporating AC and N-1 contingency constraints}, International Journal of Electrical Power \& Energy Systems 133 (2021) 107278.
\newblock \href {https://doi.org/https://doi.org/10.1016/j.ijepes.2021.107278} {\path{doi:https://doi.org/10.1016/j.ijepes.2021.107278}}.

\bibitem{QC}
C.~Coffrin, H.~L. Hijazi, P.~Van~Hentenryck, The qc relaxation: A theoretical and computational study on optimal power flow, IEEE Transactions on Power Systems 31~(4) (2016) 3008--3018.
\newblock \href {https://doi.org/10.1109/TPWRS.2015.2463111} {\path{doi:10.1109/TPWRS.2015.2463111}}.

\bibitem{OTS_QC}
K.~Bestuzheva, H.~Hijazi, C.~Coffrin, {Convex Relaxations for Quadratic On/Off Constraints and Applications to Optimal Transmission Switching}, INFORMS Journal on Computing 32~(3) (2020).

\bibitem{Li2019IEEEAccess}
Y.~Li, B.~Hu, K.~Xie, L.~Wang, Y.~Xiang, R.~Xiao, D.~Kong, Day-ahead scheduling of power system incorporating network topology optimization and dynamic thermal rating, IEEE Access 7 (2019) 35287--35301.
\newblock \href {https://doi.org/10.1109/ACCESS.2019.2904877} {\path{doi:10.1109/ACCESS.2019.2904877}}.

\bibitem{Numan_2020}
M.~Numan, D.~Feng, F.~Abbas, U.~Rahman, W.~A. Wattoo, Impact assessment of a co-optimized dynamic line rating and transmission switching topology on network expansion planning, International Transactions on Electrical Energy Systems 30~(8) (2020) e12457.
\newblock \href {https://doi.org/https://doi.org/10.1002/2050-7038.12457} {\path{doi:https://doi.org/10.1002/2050-7038.12457}}.

\bibitem{LAI_2022}
C.-M. Lai, J.~Teh, Network topology optimisation based on dynamic thermal rating and battery storage systems for improved wind penetration and reliability, Applied Energy 305 (2022) 117837.
\newblock \href {https://doi.org/https://doi.org/10.1016/j.apenergy.2021.117837} {\path{doi:https://doi.org/10.1016/j.apenergy.2021.117837}}.

\bibitem{KAWASAKI2022}
Y.~Kawasaki, N.~Yamaguchi, Unit commitment with network topology optimization considering flexibility, in: 2022 IEEE International Conference on Power Systems Technology (POWERCON), 2022, pp. 1--6.
\newblock \href {https://doi.org/10.1109/POWERCON53406.2022.9929680} {\path{doi:10.1109/POWERCON53406.2022.9929680}}.

\bibitem{KAWASAKI2024}
Y.~Kawasaki, N.~Yamaguchi, Reduction method for short-circuit capacity by unit commitment with network topology optimization, Sustainable Energy, Grids and Networks 38 (2024) 101398.
\newblock \href {https://doi.org/https://doi.org/10.1016/j.segan.2024.101398} {\path{doi:https://doi.org/10.1016/j.segan.2024.101398}}.

\bibitem{RUIZ_2009}
P.~A. Ruiz, C.~R. Philbrick, P.~W. Sauer, Wind power day-ahead uncertainty management through stochastic unit commitment policies, in: 2009 IEEE/PES Power Systems Conference and Exposition, 2009, pp. 1--9.
\newblock \href {https://doi.org/10.1109/PSCE.2009.4840133} {\path{doi:10.1109/PSCE.2009.4840133}}.

\bibitem{Papavasiliou_2013}
A.~Papavasiliou, S.~S. Oren, Multiarea stochastic unit commitment for high wind penetration in a transmission constrained network, IEEE Transactions on Sustainable Energy 61~(3) (2013) 578--592.
\newblock \href {https://doi.org/10.1287/opre.2013.1174} {\path{doi:10.1287/opre.2013.1174}}.

\bibitem{Takriti_1996}
S.~Takriti, J.~Birge, E.~Long, A stochastic model for the unit commitment problem, IEEE Transactions on Power Systems 11~(3) (1996) 1497--1508.
\newblock \href {https://doi.org/10.1109/59.535691} {\path{doi:10.1109/59.535691}}.

\bibitem{Zheng_2015}
Q.~P. Zheng, J.~Wang, A.~L. Liu, Stochastic optimization for unit commitment—a review, IEEE Transactions on Power Systems 30~(4) (2015) 1913--1924.
\newblock \href {https://doi.org/10.1109/TPWRS.2014.2355204} {\path{doi:10.1109/TPWRS.2014.2355204}}.

\bibitem{Carpentier_1996}
P.~Carpentier, G.~Gohen, J.-C. Culioli, A.~Renaud, Stochastic optimization of unit commitment: a new decomposition framework, IEEE Transactions on Power Systems 11~(2) (1996) 1067--1073.
\newblock \href {https://doi.org/10.1109/59.496196} {\path{doi:10.1109/59.496196}}.

\bibitem{Ozturk_2004}
U.~Ozturk, M.~Mazumdar, B.~Norman, A solution to the stochastic unit commitment problem using chance constrained programming, IEEE Transactions on Power Systems 19~(3) (2004) 1589--1598.
\newblock \href {https://doi.org/10.1109/TPWRS.2004.831651} {\path{doi:10.1109/TPWRS.2004.831651}}.

\bibitem{Qiu_2015}
F.~Qiu, J.~Wang, Chance-constrained transmission switching with guaranteed wind power utilization, IEEE Transactions on Power Systems 30~(3) (2015) 1270--1278.
\newblock \href {https://doi.org/10.1109/TPWRS.2014.2346987} {\path{doi:10.1109/TPWRS.2014.2346987}}.

\bibitem{Han_2023}
T.~Han, D.~J. Hill, Y.~Song, Optimal transmission switching with uncertainties from both renewable energy and n-k contingencies, IEEE Transactions on Sustainable Energy 14~(4) (2023) 1964--1978.
\newblock \href {https://doi.org/10.1109/TSTE.2023.3278072} {\path{doi:10.1109/TSTE.2023.3278072}}.

\bibitem{LPAC}
C.~Coffrin, P.~{Van Hentenryck}, A linear-programming approximation of ac power flows, https://arxiv.org/abs/1206.3614 (2013).
\newblock \href {http://arxiv.org/abs/1206.3614} {\path{arXiv:1206.3614}}.

\bibitem{PowerModels2018}
C.~Coffrin, R.~Bent, K.~Sundar, Y.~Ng, M.~Lubin, Powermodels.jl: An open-source framework for exploring power flow formulations, in: Power Systems Computation Conference, 2018, pp. 1--8.
\newblock \href {https://doi.org/10.23919/Power Systems Computation Conference.2018.8442948} {\path{doi:10.23919/Power Systems Computation Conference.2018.8442948}}.

\bibitem{ergun_optimal_2019}
H.~Ergun, J.~Dave, D.~Van~Hertem, F.~Geth, Optimal power flow for {AC–DC} grids: Formulation, convex relaxation, linear approximation, and implementation, IEEE Transactions on Power Systems 34~(4) (2019) 2980--2990.
\newblock \href {https://doi.org/10.1109/TPWRS.2019.2897835} {\path{doi:10.1109/TPWRS.2019.2897835}}.

\bibitem{Heitsch_2003}
H.~Heitsch, W.~Römisch, Scenario reduction algorithms in stochastic programming, Computational Optimization and Applications 24 (2003) 187--206.
\newblock \href {https://doi.org/https://doi.org/10.1023/A:1021805924152} {\path{doi:https://doi.org/10.1023/A:1021805924152}}.

\bibitem{Morales_2009}
J.~M. Morales, A.~J. Conejo, J.~Perez-Ruiz, Economic valuation of reserves in power systems with high penetration of wind power, in: 2009 IEEE Power \& Energy Society General Meeting, 2009.

\bibitem{Birge_2011}
J.~R. Birge, F.~Louveaux, Introduction to Stochastic Programming, 2011.
\newblock \href {https://doi.org/https://doi.org/10.1007/978-1-4614-0237-4} {\path{doi:https://doi.org/10.1007/978-1-4614-0237-4}}.

\bibitem{Papavasiliou_2011}
A.~Papavasiliou, S.~S. Oren, R.~P. O'Neill, Reserve requirements for wind power integration: A scenario-based stochastic programming framework, IEEE Transactions on Power Systems 26~(4) (2011) 2197--2206.
\newblock \href {https://doi.org/10.1109/TPWRS.2011.2121095} {\path{doi:10.1109/TPWRS.2011.2121095}}.

\bibitem{Devroye2001}
L.~Devroye, G.~Lugosi, The Kernel Density Estimate, Springer New York, New York, NY, 2001, pp. 79--97.
\newblock \href {https://doi.org/10.1007/978-1-4613-0125-7_9} {\path{doi:10.1007/978-1-4613-0125-7_9}}.

\bibitem{Jin2010}
X.~Jin, J.~Han, K-Means Clustering, Springer US, Boston, MA, 2010, pp. 563--564.
\newblock \href {https://doi.org/10.1007/978-0-387-30164-8_425} {\path{doi:10.1007/978-0-387-30164-8_425}}.

\bibitem{pglib}
S.~Babaeinejadsarookolaee, {et al.}, The power grid library for benchmarking ac optimal power flow algorithms, https://arxiv.org/abs/1908.02788 (2021).
\newblock \href {http://arxiv.org/abs/1908.02788} {\path{arXiv:1908.02788}}.

\bibitem{Gurobi}
Gurobi, Gurobi optimizer reference manual v9.3, https://www.gurobi.com/documentation/9.3/refman/index.html.

\bibitem{ipopt}
A.~{W\"{a}chter}, L.~{Biegler}, On the implementation of an interior-point filter line-search algorithm for large-scale nonlinear programming, Mathematical Programming 106~(1) (2006) 25--57.

\end{thebibliography}

\newpage
\section*{A. Appendix} \label{sec:appendix}

\subsection*{Simulation time}
\subsubsection*{30-bus test case}
\begin{table*}[h!]
\caption{Overview of the average, minimum, and maximum simulation times for the LPAC-BS model for each simulation and time series in the 30-bus test case, 24 hours and 14 days simulations. The AC-feasibility checks and redispatch have negligible simulation times compared to the LPAC-BS model. Maximum simulation time of 8 hours with a MIP gap of 0.1\%.}
\centering
{\fontsize{7pt}{10pt}\selectfont
\begin{tabular}{c|ccc|ccc|ccc}
\hline
 Time series $\rightarrow$ & \multicolumn{3}{c|}{Measured \& Forecasted}  & \multicolumn{3}{c|}{6 $\scenarios$} & \multicolumn{3}{c}{8 $\scenarios$} \\
\cline{1-1}
\multirow{2}{*}{Simulation $\downarrow$}  & \multicolumn{3}{c|}{Simulation time [s]} &  \multicolumn{3}{c|}{Simulation time [s]} & \multicolumn{3}{c}{Simulation time [s]}  \\
  & Average & Minimum & Maximum & Average & Minimum & Maximum & Average & Minimum & Maximum  \\
\hline
Hourly BS (hour)	 & 0.35 & 0.18 & 0.67 & 78.61 & 0.84 & 326.39 & 111.58 & 1.05 & 2445.99 \\
One topology (day)   & 526.62 & 51.62 & 773.76   & 4579.40 & 814.11 & 8559.84 & 6655.14 & 616.20 & 12856.59 \\
Up to one    & \multirow{2}{*}{906.22}  & \multirow{2}{*}{97.97} 	 & \multirow{2}{*}{2574.57} & \multirow{2}{*}{4588.12}  & \multirow{2}{*}{821.89} 	 & \multirow{2}{*}{7062.78} & \multirow{2}{*}{6550.26}  & \multirow{2}{*}{613.64} 	 & \multirow{2}{*}{12589.27} \\
switching action (day) &  &   & &  &   & &  &     	 \\
Up to two 	& \multirow{2}{*}{17488.02}  & \multirow{2}{*}{96.13} & \multirow{2}{*}{25969.73} & \multirow{2}{*}{4610.03}  & \multirow{2}{*}{812.73} 	 & \multirow{2}{*}{7710.44} & \multirow{2}{*}{6451.53}  & \multirow{2}{*}{628.96} & \multirow{2}{*}{11334.51} \\
switching actions (day) &  &   & &  &   & &  &    \\
\hline
\end{tabular}
}
\label{table:time_30_bus}
\end{table*}

\subsubsection*{50-bus test case}


\begin{table*}[h!]
\caption{Overview of the average, minimum, and maximum simulation times for the LPAC-BS model for each simulation and time series in the 50-bus test case, 14 days simulation. The AC-feasibility checks and redispatch have negligible simulation times compared to the LPAC-BS model. Maximum simulation time of 8 hours with a MIP gap of 0.1\%.}
\centering
{\fontsize{7pt}{10pt}\selectfont
\begin{tabular}{c|ccc|ccc|ccc}
\hline
 Time series $\rightarrow$ & \multicolumn{3}{c|}{Measured \& Forecasted}  & \multicolumn{3}{c|}{6 $\scenarios$} & \multicolumn{3}{c}{8 $\scenarios$} \\
\cline{1-1}
\multirow{2}{*}{Simulation $\downarrow$}  & \multicolumn{3}{c|}{Simulation time [s]} &  \multicolumn{3}{c|}{Simulation time [s]} & \multicolumn{3}{c}{Simulation time [s]}  \\
  & Average & Minimum & Maximum & Average & Minimum & Maximum & Average & Minimum & Maximum  \\
\hline
Hourly BS (hour)	 & 0.48 & 0.24 & 1.50 & 48.04 & 28.10 & 72.33 & 70.17 & 32.96 & 109.17 \\
One topology (day) & 269.54 & 187.58 & 420.30 & 3452.53 & 1163.80 & 5359.14 & 5804.05 & 1906.75 & 10800.16 \\
Up to one    & \multirow{2}{*}{561.96}  & \multirow{2}{*}{283.13} 	 & \multirow{2}{*}{887.11} & \multirow{2}{*}{4478.78}  & \multirow{2}{*}{1293.71} 	 & \multirow{2}{*}{7801.02} & \multirow{2}{*}{4713.23}  & \multirow{2}{*}{1879.36} 	 & \multirow{2}{*}{7792.41} \\
switching action (day) &  &   & &  &   & &  &     	 \\
Up to two 	& \multirow{2}{*}{10397.7}  & \multirow{2}{*}{2166.34} & \multirow{2}{*}{19380.95} & \multirow{2}{*}{3848.21}  & \multirow{2}{*}{1461.65} 	 & \multirow{2}{*}{6554.13} & \multirow{2}{*}{5383.03}  & \multirow{2}{*}{1844.19} 	 & \multirow{2}{*}{10800.28} \\
switching actions (day) &  &   & &  &   & &  &    \\
\hline
\end{tabular}
}
\label{table:time_50_bus}
\end{table*}
\end{document}